\documentclass[aps,prl,twocolumn,superscriptaddress,amsmath,amssymb]{revtex4-2}

\usepackage{graphicx} 
\usepackage{epstopdf} 
\usepackage{dcolumn} 
\usepackage{xcolor} 
\usepackage{amssymb} 
\usepackage{hyperref} 
\usepackage[utf8]{inputenc} 
\usepackage[english]{babel} 
\usepackage[displaymath, mathlines]{lineno} 
\usepackage{blindtext} 
\usepackage{bashful}
\usepackage{siunitx}
\usepackage[italic]{hepnames}
\usepackage{caption}
\usepackage{subcaption}
\usepackage{tikz}
\usepackage{tabularx}
\usepackage{multirow}
\usepackage{booktabs}
\usepackage{mathtools}  
\usepackage{orcidlink} 
\usepackage{soul} 

\RequirePackage{xspace}

\def\belletwo {Belle~II }

\def\epem       {\ensuremath{e^+e^-}\xspace}

\def\qqbar {\ensuremath{q\overline q}\xspace}

\def\Kbar  {\kern 0.2em\overline{\kern -0.2em K}{}\xspace}

\def\Kz    {\ensuremath{K^0}\xspace}
\def\Kzb   {\ensuremath{\Kbar^0}\xspace}
\def\KzKzb {\ensuremath{\Kz \kern -0.16em \Kzb}\xspace}
\def\Kp    {\ensuremath{K^+}\xspace}
\def\Km    {\ensuremath{K^-}\xspace}

\def\KpKm  {\ensuremath{\Kp \kern -0.16em \Km}\xspace}

\def\KL    {\ensuremath{K^0_{\scriptscriptstyle L}}\xspace}

\def\Dbar    {\kern 0.2em\overline{\kern -0.2em D}{}\xspace}

\def\Dz      {\ensuremath{D^0}\xspace}
\def\Dzb     {\ensuremath{\Dbar^0}\xspace}
\def\DzDzb   {\ensuremath{\Dz {\kern -0.16em \Dzb}}\xspace}
\def\Dp      {\ensuremath{D^+}\xspace}
\def\Dm      {\ensuremath{D^-}\xspace}

\def\DpDm    {\ensuremath{\Dp {\kern -0.16em \Dm}}\xspace}
\def\Dstar   {\ensuremath{D^*}\xspace}

\def\Dstst {\ensuremath{D^{**}}\xspace}
\def\Dzstar  {\ensuremath{D_{0}^{*}}\xspace}
\def\Done  {\ensuremath{D_{1}}\xspace}
\def\Doneprime  {\ensuremath{D'_{1}}\xspace}
\def\Dtwostar  {\ensuremath{D_{2}^{*}}\xspace}

\def\B       {\ensuremath{B}\xspace}
\def\Bbar    {\kern 0.18em\overline{\kern -0.18em B}{}\xspace}

\def\BB      {\ensuremath{B\Bbar}\xspace}
\def\Bz      {\ensuremath{B^0}\xspace}
\def\Bzb     {\ensuremath{\Bbar^0}\xspace}
\def\BzBzb   {\ensuremath{\Bz {\kern -0.16em \Bzb}}\xspace}
\def\Bu      {\ensuremath{B^+}\xspace}
\def\Bub     {\ensuremath{B^-}\xspace}

\def\BpBm    {\ensuremath{\Bu {\kern -0.16em \Bub}}\xspace}

\def\jpsi     {\ensuremath{{J\mskip -3mu/\mskip -2mu\psi\mskip 2mu}}\xspace}

\mathchardef\Upsilon="7107
\def\Y#1S{\ensuremath{\Upsilon{(#1S)}}\xspace}

\def\FourS {\Y4S}

\mathchardef\Deltares="7101
\mathchardef\Xi="7104
\mathchardef\Lambda="7103
\mathchardef\Sigma="7106
\mathchardef\Omega="710A

\def\Deltabar{\kern 0.25em\overline{\kern -0.25em \Deltares}{}\xspace}
\def\Lbar{\kern 0.2em\overline{\kern -0.2em\Lambda\kern 0.05em}\kern-0.05em{}\xspace}
\def\Sigbar{\kern 0.2em\overline{\kern -0.2em \Sigma}{}\xspace}
\def\Xibar{\kern 0.2em\overline{\kern -0.2em \Xi}{}\xspace}
\def\Obar{\kern 0.2em\overline{\kern -0.2em \Omega}{}\xspace}
\def\Nbar{\kern 0.2em\overline{\kern -0.2em N}{}\xspace}
\def\Xb{\kern 0.2em\overline{\kern -0.2em X}{}\xspace}

\newcommand{\tev}{\ensuremath{\mathrm{\,Te\kern -0.1em V}}\xspace}
\newcommand{\gev}{\ensuremath{\mathrm{\,Ge\kern -0.1em V}}\xspace}
\newcommand{\mev}{\ensuremath{\mathrm{\,Me\kern -0.1em V}}\xspace}
\newcommand{\kev}{\ensuremath{\mathrm{\,ke\kern -0.1em V}}\xspace}
\newcommand{\ev}{\ensuremath{\mathrm{\,e\kern -0.1em V}}\xspace}
\newcommand{\gevc}{\ensuremath{{\mathrm{\,Ge\kern -0.1em V\!/}c}}\xspace}
\newcommand{\mevc}{\ensuremath{{\mathrm{\,Me\kern -0.1em V\!/}c}}\xspace}
\newcommand{\gevcc}{\ensuremath{{\mathrm{\,Ge\kern -0.1em V\!/}c^2}}\xspace}
\newcommand{\gevsqcccc}{\ensuremath{{\mathrm{\,Ge\kern -0.1em V^2\!/}c^4}}\xspace}
\newcommand{\mevcc}{\ensuremath{{\mathrm{\,Me\kern -0.1em V\!/}c^2}}\xspace}

\def\barn{\ensuremath{{\rm \,b}}\xspace}

\def\mus  {\ensuremath{\rm \,\mus}\xspace}

\def\mus        {\ensuremath{\,\mu{\rm s}}\xspace}    

\def\to                 {\ensuremath{\rightarrow}\xspace}

\def\gsim{{~\raise.15em\hbox{$>$}\kern-.85em
		\lower.35em\hbox{$\sim$}~}\xspace}
\def\lsim{{~\raise.15em\hbox{$<$}\kern-.85em
		\lower.35em\hbox{$\sim$}~}\xspace}

\newcommand{\lumion}{\ensuremath{\SI{189}{\per\femto\barn}}\xspace}
\newcommand{\lumioff}{\ensuremath{\SI{18}{\per\femto\barn}}\xspace}
\newcommand{\cms}{\ensuremath{\,*}\xspace}
\newcommand{\plB}{\ensuremath{p_{\ell}^{B}}\xspace}

\newcommand*{\pllab}{\ensuremath{p_{\ell}}\xspace}
\newcommand*{\pelab}{\ensuremath{p_{e}}\xspace}
\newcommand*{\pmulab}{\ensuremath{p_{\mu}}\xspace}
\newcommand*{\thetallab}{\ensuremath{\theta_{\ell}}\xspace}

\newcommand*{\pmuTnlab}{\ensuremath{p_{\mu}}\xspace}
\newcommand*{\mmsq}{\ensuremath{M_\mathrm{miss}^2}\xspace}
\newcommand*{\mx}{\ensuremath{M_X}\xspace}
\newcommand*{\qsq}{\ensuremath{q^2}\xspace}

\newcommand*{\btag}{\ensuremath{B_\text{tag}}\xspace}
\newcommand*{\bsig}{\ensuremath{B_\text{sig}}\xspace}

\newcommand{\RX}{\ensuremath{R(X_{\tau/\ell})}\xspace}
\newcommand{\RXe}{\ensuremath{R(X_{\tau/e})}\xspace}
\newcommand{\RXmu}{\ensuremath{R(X_{\tau/\mu})}\xspace}

\newcommand{\RXc}{\ensuremath{R(X_c)}\xspace}
\newcommand{\RXu}{\ensuremath{R(X_u)}\xspace}
\newcommand{\RD}{\ensuremath{R(D)}\xspace}
\newcommand{\RDst}{\ensuremath{R(D^{*})}\xspace}
\newcommand{\RDorDst}{\ensuremath{R(D^{(*)})}\xspace}
\newcommand{\RDstst}{\ensuremath{R(D^{**})}\xspace}

\newcommand{\xtaunu}{\ensuremath{X \tau \nu}\xspace}

\newcommand{\xctaulnu}{\ensuremath{X_c \tau(\ell) \nu}\xspace}

\newcommand{\xlnu}{\ensuremath{X \ell \nu}\xspace}
\newcommand{\xclnu}{\ensuremath{X_c \ell \nu}\xspace}
\newcommand{\xulnu}{\ensuremath{X_u \ell \nu}\xspace}
\newcommand{\bxtaulnu}{\ensuremath{B\to X \tau(\ell) \nu}\xspace}
\newcommand{\bxtaunu}{\ensuremath{B\to \xtaunu}\xspace}

\newcommand{\bxlnu}{\ensuremath{B\to \xlnu}\xspace}
\newcommand{\bxclnu}{\ensuremath{B\to \xclnu}\xspace}
\newcommand{\bxulnu}{\ensuremath{B\to \xulnu}\xspace}

\newcommand{\bclnu}{\ensuremath{b \to c \ell \nu}\xspace}

\newcommand{\bctaunu}{\ensuremath{b \to c \tau \nu}\xspace}
\newcommand{\butaunu}{\ensuremath{b \to u \tau \nu}\xspace}

\newcommand{\butaulnu}{\ensuremath{b \to u \tau(\ell) \nu}\xspace}
\newcommand{\bdtaulnu}{\ensuremath{B \to D \tau(\ell) \nu}\xspace}
\newcommand{\bdtaunu}{\ensuremath{B \to D \tau \nu}\xspace}
\newcommand{\bdlnu}{\ensuremath{B \to D \ell \nu}\xspace}
\newcommand{\bdsttaulnu}{\ensuremath{B \to D^* \tau(\ell) \nu}\xspace}
\newcommand{\bdsttaunu}{\ensuremath{B \to D^* \tau \nu}\xspace}
\newcommand{\bdstlnu}{\ensuremath{B \to D^* \ell \nu}\xspace}
\newcommand{\bdordsttaulnu}{\ensuremath{B \to D^{(*)} \tau(\ell) \nu}\xspace}
\newcommand{\bddstordststtaulnu}{\ensuremath{B \to D^{(*, **)} \tau(\ell) \nu}\xspace}
\newcommand{\bdordsttaunu}{\ensuremath{B \to D^{(*)} \tau \nu}\xspace}
\newcommand{\bdordstlnu}{\ensuremath{B \to D^{(*)} \ell \nu}\xspace}
\newcommand{\bdststtaulnu}{\ensuremath{B \to D^{**} \tau(\ell) \nu}\xspace}

\newcommand{\dordstpipitaulnu}{\ensuremath{D^{(*)} \pi \pi \tau(\ell) \nu}\xspace}

\newcommand{\dordstetataulnu}{\ensuremath{D^{(*)} \eta \tau(\ell)\nu}\xspace}

\newcommand{\bdordstetalnu}{\ensuremath{B \to D^{(*)} \eta \ell \nu}\xspace}

\newcommand{\BFxtaunu}{\ensuremath{\mathcal{B}(\bxtaunu)}\xspace}
\newcommand{\BFdtaunu}{\ensuremath{\mathcal{B}(\bdtaunu)}\xspace}
\newcommand{\BFdsttaunu}{\ensuremath{\mathcal{B}(\bdsttaunu)}\xspace}

\newcommand{\BFdgapxutaunu}{\ensuremath{\mathcal{B}(B \to D^{**}_\text{(gap)}/X_u \tau \nu)}\xspace}
\newcommand{\BFxlnu}{\ensuremath{\mathcal{B}(\bxlnu)}\xspace}

\newcommand{\BFdlnu}{\ensuremath{\mathcal{B}(\bdlnu)}\xspace}
\newcommand{\BFdstlnu}{\ensuremath{\mathcal{B}(\bdstlnu)}\xspace}
\newcommand{\BFtaulnunu}{\ensuremath{\mathcal{B}(\tau \to \ell \nu \nu)}\xspace}

\def\jpsill         {\ensuremath{\jpsi \to \ell^+\ell^-}\xspace}

\begin{document}

\pacs{\input{pacs}}

\title{First Measurement of \RX as an Inclusive Test of the \bctaunu Anomaly}

\author{I.~Adachi\,\orcidlink{0000-0003-2287-0173}} 
\author{K.~Adamczyk\,\orcidlink{0000-0001-6208-0876}} 
\author{L.~Aggarwal\,\orcidlink{0000-0002-0909-7537}} 
\author{H.~Ahmed\,\orcidlink{0000-0003-3976-7498}} 
\author{H.~Aihara\,\orcidlink{0000-0002-1907-5964}} 
\author{N.~Akopov\,\orcidlink{0000-0002-4425-2096}} 
\author{A.~Aloisio\,\orcidlink{0000-0002-3883-6693}} 
\author{D.~M.~Asner\,\orcidlink{0000-0002-1586-5790}} 
\author{H.~Atmacan\,\orcidlink{0000-0003-2435-501X}} 
\author{T.~Aushev\,\orcidlink{0000-0002-6347-7055}} 
% ...
\collaboration{The Belle II Collaboration}

\begin{abstract}
We measure the tau-to-light-lepton ratio of inclusive $B$-meson branching fractions
$\RX \equiv \BFxtaunu/\BFxlnu$, where $\ell$ indicates an electron or muon, and thereby test the universality of charged-current weak interactions.
We select events that have one fully reconstructed \B meson and a charged lepton candidate from \lumion of electron-positron collision data collected with the Belle II detector. We find $\RX = 0.228 \pm 0.016~(\mathrm{stat}) \pm 0.036~(\mathrm{syst})$, in agreement with standard-model expectations. This is the first direct measurement of \RX.
\end{abstract}

\maketitle


In the standard model, all leptons share the same electroweak coupling, a symmetry known as lepton universality. Semileptonic decays of \B mesons into a charmed hadronic state, a lepton, and a neutrino provide excellent sensitivity to potential lepton-universality-violating (LUV) new interactions~\cite{bernlochner_review}.
The combination of experimental LUV tests in the rate of such exclusive decays to $D^{(*)}$ mesons and $\tau$ leptons relative to the light leptons, $\ell = e$ or $\mu$, from the BaBar~\cite{babar_1, babar_2}, Belle~\cite{belle_hadronic, belle_pol_PRL, belle_pol_PRD, belle_semileptonic}, Belle~II~\cite{b2_rdst}, and LHCb~\cite{lhcb_new1, lhcb_new2} experiments has a $3.3\sigma$ tension with the standard model expectation~\cite{hflav2023}. This could indicate an enhanced coupling of the $b$ quark to the $\tau$ lepton, as predicted in some beyond-standard-model scenarios~\cite{twoHDM, SLQ}. 
Measurements of the inclusive $b$-hadron branching fraction $\mathcal{B}(b\text{-admix}\to \xtaunu)$ at the LEP experiments~\cite{aleph, delphi, l3_1, l3_2, opal}, however, are consistent with the standard model. Here, $X$ indicates the generic hadronic final state that originates from \bctaunu or, rarely, \butaunu decays.

We present a complementary probe of LUV through the first measurement of the tau-to-light-lepton ratio of inclusive semileptonic \B-meson branching fractions, $\RX \equiv \BFxtaunu/\BFxlnu$. This approach incorporates $D$ and \Dstar mesons regardless of their decay mode and includes a $14\% - 20\%$ expected contribution from unexplored semitauonic \B-meson decays~\cite{supplemental}\nocite{mannel_rxc}. The predictions for \RX and those for the ratios of exclusive decays \RDorDst are based on different theoretical input~\cite{kvos, ligeti_rxc}. Consequently, this measurement provides a lepton-universality test that is statistically and theoretically distinct from \RDorDst, sensitive to different systematic uncertainties, and is potentially more precise~\cite{snowmass, bernlochner_review}.

We use electron-positron collision data collected by the \belletwo experiment between 2019 and 2021 at the center-of-mass energy $\sqrt{s}=10.58$~GeV, corresponding to the \FourS resonance, which decays almost exclusively into \BzBzb or \BpBm. The integrated luminosity is \lumion, equivalent to approximately $198\times 10^6$ \BB pairs. We use an additional off-resonance dataset (\lumioff) collected $\SI{60}{\MeV}$ below the energy of the \FourS resonance to determine the expected backgrounds from continuum processes $\epem\to\qqbar$, where $q$ indicates $u, d, s$, or $c$ quarks.

We reconstruct one \B meson in a fully hadronic decay mode (the partner \B, labeled \btag) and associate remaining particle candidates, which must include a lepton identified as an electron or muon, with the accompanying \B meson (the signal, \bsig). The event is called signal if the lepton is the decay product of a primary $\tau$ lepton in a \bxtaunu decay and normalization if the lepton is the primary lepton in a \bxlnu decay. We simultaneously extract the signal and normalization yields with a two-dimensional fit to the distribution of \plB, the lepton momentum in the rest frame of the \bsig meson, and the missing mass squared $\mmsq=[(\sqrt{s}, \vec{0}) - P_{\btag}^{\cms} - P_{X}^{\cms} - P_{\ell}^{\cms}]^2$, where  $P_{\btag}^{\cms}$, $P_{X}^{\cms}$, and $P_{\ell}^{\cms}$ are the measured four-momenta of the partner \B meson, the hadronic system $X$, and the signal-lepton candidate, in the center-of-mass frame, respectively.


The \belletwo detector at the SuperKEKB asymmetric-energy electron-positron collider~\cite{superkekb} consists of a nearly hermetic solenoidal magnetic spectrometer surrounded by particle-identification, electromagnetic-calorimetry (ECL), and muon subdetectors~\cite{b2tdr, b2tip}. The detector has a cylindrical barrel region that is nearly coaxial with the beams and closed on either end by forward and backward end caps. 


We use simulation to model the signal, normalization, and backgrounds, identify selection criteria, and to calculate reconstruction efficiencies. The software packages \textsc{\small{EvtGen}}~\cite{evtgen}, \textsc{\footnotesize{PYTHIA}}~\cite{pythia8}, and \textsc{\footnotesize{KKMC}}~\cite{kkmc} are used to model particle production and decay, \textsc{\small{Photos}}~\cite{PHOTOS} for photon radiation from charged particles, and \textsc{\small{Geant4}}~\cite{geant4} for material interaction and detector response. Simulated beam-induced backgrounds are overlaid on the events~\cite{BeamBKG}. Simulated events are processed as collision data with the Belle~II analysis software {\fontfamily{qcr}\selectfont basf2}~\cite{basf2, basf2-zenodo}. Simulated $\epem \to \FourS \to \BB$ samples, equivalent to an integrated luminosity of $\SI{900}{\per\femto\barn}$, contain known semileptonic and hadronic \B decays and additional hadronic \B decays modeled using \textsc{\footnotesize{PYTHIA}}. The signal (normalization) model includes the following exclusive decays, with charge conjugation implied throughout: \bdtaulnu, \bdsttaulnu, and \bdststtaulnu, where \Dstst collectively indicates the excited charmed states \Dzstar, \Doneprime, \Done, and \Dtwostar, whose masses and widths are taken from Ref.~\cite{pdg}. The \bdordsttaulnu decays are modeled with the BLPRXP form-factor parametrization~\cite{blprxp}. The modeling of \bdststtaulnu decays is based on the BLR model~\cite{bernlochner_dstst, bernlochner_rdstst}.
Semileptonic \B decays into the nonresonant final states \dordstpipitaulnu and \dordstetataulnu are used to fill the difference between the sum of individual branching ratios of exclusive decays, \bddstordststtaulnu, and the total semileptonic \B decay widths. These ``gap modes" are included in dedicated simulated samples that use intermediate, broad \Dstst resonances and are modeled with BLR. We take the total width for decays to light leptons from Ref.~\cite{pdg}; the widths for semitauonic modes are based on standard-model predictions relative to the light-lepton decay rates~\cite{bernlochner_rd_rdst, bernlochner_rdstst, ligeti_rxc, kvos}. Semileptonic \B decays including \butaulnu quark transitions are simulated in a hybrid model~\cite{xu_hybrid} that is updated and scaled according to Refs.~\cite{pdg, xutaunu}.\par


In order to identify signal events in the data, we first use the Full Event Interpretation to reconstruct the \btag meson in a fully hadronic decay mode~\cite{fei}. This allows each final-state particle to be uniquely associated with one of the two \B mesons and increases the signal purity. 
The selection follows Ref.~\cite{Remu}, resulting in $63\%$ charged and $37\%$ neutral \btag candidates. 


We select signal-lepton candidates from charged-particle trajectories (tracks) remaining after the \btag reconstruction. We require that the lepton charge corresponds to the charge of a primary lepton from the semileptonic decay of a $B$ meson of opposite flavor to the \btag candidate. Lepton tracks are required to be within the angular acceptance of the tracking system and to point back to within $\SI{1}{\centi\meter}$ ($\SI{1.5}{\milli\meter}$) of the interaction point in three dimensions for electrons (muons).

Muon candidates are identified using a likelihood ratio that combines particle-identification information from all relevant subdetectors. In order to efficiently reject misidentified muon candidates, we require their transverse momenta to be
above $0.4\gevc$ and lab-frame momenta \pmulab to be above $0.7\gevc$. The muon identification efficiency is measured from control channels to be on average $90\%$ for $\pmulab>1\gevc$, corresponding to a muon misidentification probability for pions (kaons) of $2.5\%$ ($1\%$). For $\pmulab < 1\gevc$, the efficiency averages to $70\%$, with a pion (kaon) misidentification probability as a muon of $4\%$ ($0.2\%$).

Electron candidates are identified using a multiclass boosted-decision-tree classifier that exploits several ECL observables in combination with particle-identification likelihoods~\cite{LeptonIDBDT}. We require their transverse momenta to be
above $0.3\gevc$ and lab-frame momenta \pelab to be above $0.5\gevc$ and correct their four-momenta to recover bremsstrahlung radiation. A stringent classifier threshold results in average misidentification probabilities for pions (kaons) with $\pelab>1\gevc$ of $0.1\%$ ($0.02\%$); and for $\pelab < 1\gevc$, of less than $0.1\%$ ($0.1\%$), with electron identification efficiencies of roughly $95\%$ for $\pelab > 1\gevc$ and $70\%$ for $\pelab < 1\gevc$. 

In $3\%$ of events, two or more signal-lepton candidates pass these selections. In these events, we select the candidate with the highest identification likelihood, with 81\% efficiency for simulated signal leptons.

Backgrounds from $\Upsilon(4S)\to\BB$ decays include hadrons misidentified as leptons (fakes) and leptons originating mainly from decays of charmed hadrons (secondaries). We suppress muon fakes from pions or kaons by rejecting combinations that are consistent with $\omega \to \pi^+\pi^-\pi^0$, $K^{*0}\to\pi^-K^+$, $D^0\to K^- \pi^- \pi^+ \pi^+$, $D^+\to \pi^+\pi^+\pi^-+ [\pi^0\text{ or }\pi^+\pi^-]$, and $D^+\to K^-\pi^+\pi^+(\pi^0)$ decays, assigning the pion or kaon mass to the muon candidate. We suppress secondaries from photon conversion and $\pi^0\to e^+ e^- (\gamma)$ decays by rejecting signal-electron candidates with \bsig-frame momentum \plB below $1\gevc$ if they combine with an oppositely charged particle in a displaced vertex or have an $e^+e^-$ invariant mass below $0.15\gevcc$. We reject signal-lepton candidates that can combine with an oppositely charged lepton yielding the \jpsill mass.

Following Ref.~\cite{Remu}, we obtain corrections and uncertainties for lepton-identification performance in discrete intervals of lab-frame momentum (\pllab), angle with respect to the electron beam (\thetallab), and charge ($q$), using dedicated data samples. We suppress continuum background, described by off-resonance data, using event-topology variables, and correct its yield and the kinematic properties of the final-state particles following the same reference.


The hadronic system $X$ is reconstructed using remaining tracks and energy deposits in the ECL (clusters) that are not associated with any tracks. Clusters are required to be at least $\SI{30}{\centi\meter}$ from the nearest track and have energies greater than $40$, $55$, and $\SI{90}{\MeV}$ in the forward, barrel, and backward regions, respectively. 
Tracks are required to be consistent with originating from the interaction point and are assigned masses according to the first satisfied particle-identification criterion in the sequence electron, muon, kaon, and proton. Tracks satisfying none of these criteria are assigned the pion mass. We define \mx to be the invariant mass of all reconstructed particles associated with the $X$ system.


\begin{figure*}[htbp]
	\centering
	\includegraphics[width=0.516\linewidth]{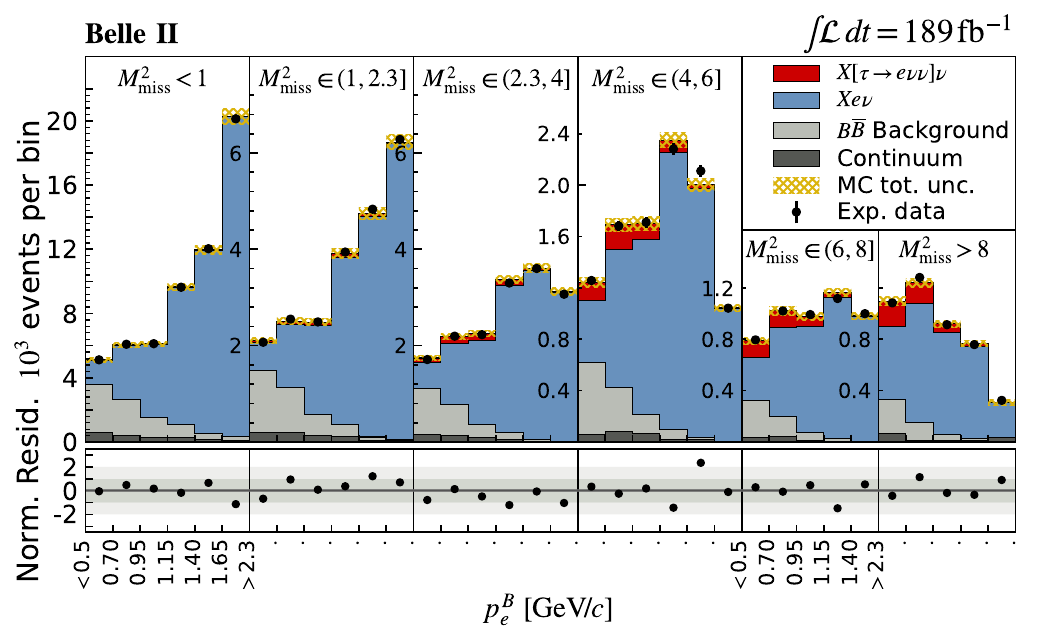}\hspace{-2.7mm}\sbox0{\includegraphics{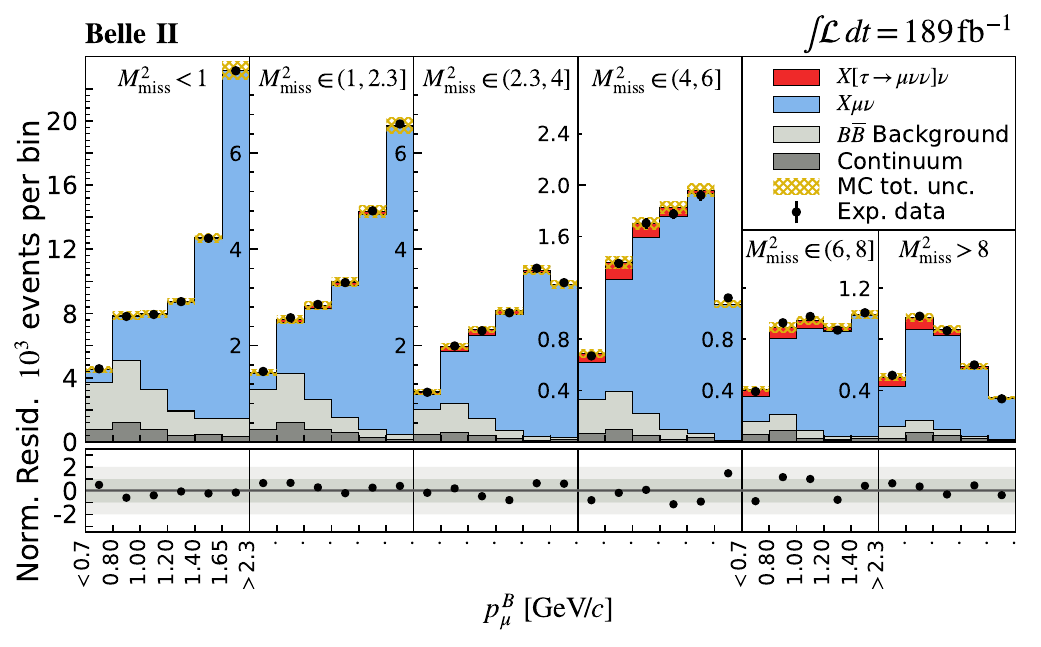}}
	\includegraphics[clip,trim={0.045\wd0} {0.0064516129\ht0} 0 0,width=0.49278\linewidth]{templates_postfit_mu.pdf}
	\caption{Two-dimensional distributions of electron (left) and muon (right) momentum in the \bsig rest frame \plB and the missing mass squared \mmsq, flattened to one dimension in intervals as used in the signal extraction fit, with the fit results overlaid. The hatched area shows the total statistical and systematic uncertainty, added in quadrature for each interval. The residuals are normalized to the statistical uncertainty of the data points and the \mmsq intervals are given in units of \gevsqcccc.}
	\label{fig:fit_templates}
\end{figure*}

In order to assess the modeling of normalization and background events, we define two samples that are depleted of signal events. The high-\plB ($\plB > 1.4\gevc$) sample consists of $95\%$ \bxlnu events. The same-flavor sample contains events where both $B$ mesons have the same flavor, and it is enriched with fakes, secondaries, and continuum ($77\%$), but also includes \bxtaulnu decays from neutral $B$-meson oscillations ($23\%$).

The simulated \mx distribution in the high-\plB sample has a significant deficit relative to the experimental distribution for low \mx values, and a significant excess for high \mx values, suggesting mismodeling of the charmed meson decays that largely constitute the $X$ system. Such mismodeling would likely introduce inconsistencies in the pion and kaon multiplicities, which we also observe. These effects cannot be attributed to the modeling of \bdordsttaulnu or \bdststtaulnu decays, or of detector response.  The disagreement is particularly sensitive to the simulated proportion of $D$ decays to \KL mesons. 

To correct for this mismodeling, we reweight the simulated \bxtaulnu events using the experimental-to-simulated yield ratio in 17 intervals of \mx using the high-\plB sample. The weights range between $0.80$ and $1.27$, with uncertainties from $0.01$ to $0.18$ that are dictated by the number of simulated and experimental events in each bin. The largest weights are measured at low \mx, a region enriched with events with at least one \KL meson. 
We measure these weights in five equally populated \plB intervals in each \mx interval, and find statistically consistent values.
In order to accommodate any \plB dependence potentially hidden by the statistical uncertainties, we extrapolate linearly to the signal-enriched region $\plB < 1.25\gevc$ and assign the absolute difference between the nominal and extrapolated weights as an additional uncertainty. We apply these factors as weights to \bxtaulnu events, then rescale the proportions of the known individual semileptonic $B$ decays so that they are preserved. 

We also reweight secondary-lepton and muon-fake events in the simulated \BB backgrounds. We derive weights for secondary leptons using the same-flavor electron sample, in which $98\%$ of the \BB backgrounds are secondaries. We determine the weights in two-dimensional intervals of \pllab and \mx. These weights are equally applicable to electron and muon secondaries, as they predominantly arise from the same $D$ decay modes. We derive weights for muon fakes from the remaining experiment-to-simulation deviations. The impact of this reweighting on the \BB background yields is treated as an uncertainty on the muon fake-to-secondary composition. Secondaries from hadronic $B$ decays to multiple charmed hadrons are assigned shape and yield uncertainties that cover up to twice the corrections derived from the same-flavor secondaries due to their low purity in this sample. 

These simulation reweightings significantly ameliorate mismodeling in kinematic variables correlated to the $X$ system, in particular \mmsq~\cite{supplemental}. 


We extract the signal and normalization yields for the electron and muon modes, $N^\text{meas}_{\tau (\ell)}$, from a simultaneous maximum-likelihood fit to the binned two-dimensional distributions of \plB and \mmsq (Fig.~\ref{fig:fit_templates}). For each lepton flavor we define signal, normalization, \BB background, and continuum components, and associate each with a histogram template and corresponding yield parameter. The signal, normalization, and \BB background yields are unconstrained. The continuum component has a Gaussian constraint on its yield derived from off-resonance data.  


The statistical and systematic uncertainties on the templates are incorporated in the likelihood definition via nuisance parameters, one for each (\plB, \mmsq) bin for each component. Constraints on the nuisance parameters are encoded in a global covariance matrix for bins and components, constructed by summing the covariance matrices of all individual uncertainty sources. 

Uncertainties on the track-reconstruction efficiency are estimated with control samples, and are propagated as a $0.3\%$ uncertainty per track in each event. The uncertainties associated with the lepton-identification-performance weights are also provided by auxiliary measurements~\cite{Remu}. We propagate these to \RX uncertainties by assuming that they are fully correlated for events of a given lepton or hadron fake type that share a (\pllab, \thetallab, $q$) bin, and are uncorrelated otherwise. 
We derive the uncertainties associated with the simulation reweighting via 500 random variations of the event weights. These uncertainties are assumed to be fully correlated for events in the same \mx (or (\pllab, \mx), for \BB background) bin, and, initially, to be uncorrelated otherwise. The total yield is fixed in each variation, thereby introducing correlations caused by bin migrations.

We incorporate branching-fraction uncertainties by deriving efficiency uncertainties and bin covariances based on histogram shape differences caused by their $\pm 1\sigma$ variations.
For \bxulnu and \bdordstlnu decays, we use the latest values~\cite{pdg, hflav}, combining the results of neutral and charged \B mesons under the assumption of isospin symmetry. For the remaining \bclnu decays, not all final states have been measured to date. We estimate the unknown branching fractions by extrapolating from existing measurements to the unobserved $D^{**}$ final-state decays, assuming isospin symmetry.
Among the nonresonant gap modes, only the decay $B\rightarrow D^{(*)} \pi^+ \pi^- \ell \nu$ is measured~\cite{dpipi_babar}. This result is extrapolated to the other charge configurations to estimate their total branching fractions. The remaining gap modes, \bdordstetalnu, are assigned a $100\%$ branching fraction uncertainty. The fit to data reduces the uncertainties on the gap-mode branching fractions by exploiting its distinctive shape. Branching fractions of semitauonic $B$ decays are derived by combining the corresponding light-lepton branching fractions with the standard-model predictions of \RDorDst, \RDstst, and \RXu~\cite{bernlochner_rd_rdst, bernlochner_rdstst, xutaunu}; semitauonic gap-mode branching fractions are based on the average of the \RX predictions of Refs.~\cite{ligeti_rxc, kvos}, with  relative fractions assumed to be equivalent to those of the corresponding light-lepton gap modes.

Decay-model parameters (form factors) are varied according to their covariances using the \textsc{\footnotesize{HAMMER}} software package~\cite{hammer}. The full differences between the BLPRXP parameter prediction and parametrizations using the BGL~\cite{BGL, DlnuBGL, DstlnuBGL} (CLN~\cite{CLN, hflav}) model are treated as additional uncertainties for \bdordstlnu (\bdordsttaunu) processes.\par

We assume that the tagging efficiency cancels in the \RX ratio. This assumption is supported by the agreement between experimental and simulated distributions of all relevant \btag quantities.

After all selections and corrections, we determine efficiencies from the ratio of the number of selected signal (normalization) events in simulation $N^{\mathrm{sel}}_{\tau \to \ell (\ell)}$ to the number generated $N^\text{gen}_{\tau \to \ell(\ell)}$. The electron-mode efficiency is $(1.50\pm 0.02)\times10^{-3}$ in the signal mode and $(2.29\pm 0.03) \times10^{-3}$ in the normalization mode, with a correlation of $0.62$. The respective muon-mode efficiencies are lower, $(1.12\pm 0.02) \times10^{-3}$ and $(2.15\pm 0.03) \times10^{-3}$, due to more-restrictive \pmuTnlab thresholds, with a correlation of $0.71$.

We fit the experimental (\plB, \mmsq) spectra as shown in Fig.~\ref{fig:fit_templates} and measure electron (muon) normalization yields of $N_{e}^{\text{meas}}=\SI{95690}{} \pm 770$ ($N_{\mu}^{\text{meas}}=\SI{89970}{}\pm 810$) and signal yields of $N_{\tau \to e}^{\text{meas}}=2590\pm 450$ ($N_{\tau \to \mu}^{\text{meas}}=1810\pm 460$), with correlations of $-0.53$ $(-0.54).$ From these yields and correlations, we calculate \RX and its uncertainty using $N_{\tau}^{\text{gen}} = N_{\tau \to \ell}^{\text{gen}}/\BFtaulnunu$ via
$\RX = (N_{\tau \to \ell}^{\text{meas}}/N_{\ell}^{\text{meas}})(N_{\ell}^{\text{sel}}/N_{\tau \to \ell}^{\text{sel}}) (N_{\tau}^{\text{gen}}/N_{\ell}^{\text{gen}})$ and the appropriate uncertainty propagation.


\begin{table}[t]
	\caption{Relative statistical and systematic uncertainties on the value of \RX for electrons, muons, and their combination ($\ell$). Detailed descriptions of each source are provided in the text.}
	\begin{tabularx}{\columnwidth}{l @{\hskip 8mm} c @{\hskip 8mm} c @{\hskip 8mm}c}
		\toprule
		\toprule
		\multirow{2}{*}{Source}     & \multicolumn{3}{c}{Uncertainty [\%]} \\ 
		& $e$ & $\mu$ & $\ell$ \\
		\midrule
		Experimental sample size               & 8.8 & 12.0 & 7.1 \\
		Simulation sample size         & 6.7 & 10.6 & 5.7 \\
		Tracking efficiency            & 2.9 &  3.3 & 3.0 \\
		Lepton identification          & 2.8 &  5.2 & 2.4 \\
		\xclnu reweighting & 7.3 &  6.8 & 7.1 \\
		\BB background reweighting  & 5.8 & 11.5 & 5.7 \\
		\xlnu branching fractions       & 7.0 &  10.0 & 7.7 \\
		\xtaunu branching fractions     & 1.0 &  1.0 & 1.0 \\
		\xctaulnu form factors         & 7.4 &  8.9 & 7.8 \\
		\midrule
		Total                          & 18.1 & 25.6 & 17.3 \\
		\bottomrule
		\bottomrule
	\end{tabularx}
	\label{tab:uncertainties}
\end{table}

We estimate the size of each systematic uncertainty by refitting the simulated spectrum with all systematic sources fixed and then with all but one source fixed, and take the quadrature difference between the two. 

The resulting uncertainties are summarized in Table~\ref{tab:uncertainties}. The largest uncertainties are associated with the experimental and simulation sample sizes.
Normalization and \BB background shape uncertainties associated with the simulation reweightings are driven by the sample sizes of the control samples. They should decrease with larger sample sizes like statistical uncertainties, as should the branching-fraction uncertainties, which are dominated by constraints on the $100\%$ uncertainty assigned to the branching fraction of the nonresonant gap modes from the fit to data. These sources are comparable to the form-factor uncertainties, which are dominated by deviations between form-factor model parametrizations for \bdstlnu processes.


We find \RX for electrons and muons of
\begin{align}
	\RXe &= 0.232 \pm 0.020~(\mathrm{stat}) \pm 0.037~(\mathrm{syst}) \text{, and} \notag  \\ 
	\RXmu &= 0.222 \pm 0.027~(\mathrm{stat}) \pm 0.050~(\mathrm{syst}) \notag \text{,}
\end{align}
respectively. By combining light-lepton flavors in a weighted average of correlated values, we find
\begin{equation}\notag
	\RX = 0.228 \pm 0.016~(\mathrm{stat}) \pm 0.036~(\mathrm{syst}) \text{.}
\end{equation}

This work started as a blind analysis. Unblinding of an earlier version exposed a significant correlation of the results with the lepton momentum threshold, attributed to a biased selection applied in an early data-processing step and to insufficient treatment of low-momentum backgrounds. We reblinded, removed the problematic selection, tightened lepton requirements, and introduced the lepton-secondary and muon-fake reweightings. The results are now independent of the lepton momentum threshold, and are consistent between subsets of the full dataset when split by lepton charge, tag flavor, lepton polar angle, and data collection period. We verify that the reweighting uncertainties cover mismodeling of $D$-meson decays by varying the branching ratio of each decay $D\to K(\mathrm{anything})$ within its uncertainty as provided in Ref.~\cite{pdg} while fixing the total event normalization.

\begin{figure}[t]
	\centering
	\includegraphics[width=\linewidth]{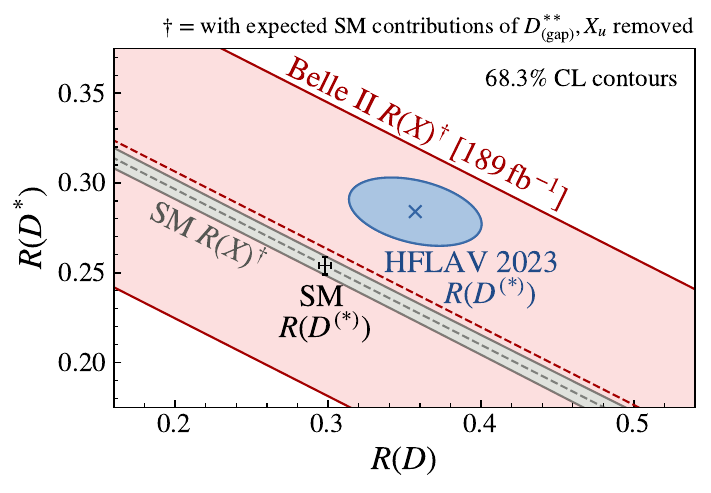}
	\caption{Constraints on \RDorDst from the measured \RX value (red), compared to the world average of \RDorDst (blue~\cite{hflav2023}) and the standard model expectation (gray and black~\cite{hflav, hflav2023}). We describe the calculation of the constraining $\RX^\dagger$ in the Supplemental Material~\cite{supplemental}.
	}
	\label{fig:result_plot}
\end{figure}

Our result is in agreement with an average of standard-model predictions of $0.223 \pm 0.005$~\cite{ligeti_rxc, kvos, xutaunu} but also is consistent with a hypothetically enhanced semitauonic branching fraction as indicated by the \RDorDst world averages~\cite{hflav} (cf. Fig.~\ref{fig:result_plot}). Because of distinct experimental strategies and small statistical overlap (approximately $0.4\%$ shared events), the total correlation between this measurement and the exclusive measurement of \RDst in Ref.~\cite{b2_rdst} is estimated to be below $0.1$. Therefore, \RX is a largely independent probe of the \bctaunu anomaly.\newline
\par

This work, based on data collected using the Belle II detector, which was built and commissioned prior to March 2019,
was supported by
Higher Education and Science Committee of the Republic of Armenia Grant No.~23LCG-1C011;
Australian Research Council and Research Grants
No.~DP200101792, 
No.~DP210101900, 
No.~DP210102831, 
No.~DE220100462, 
No.~LE210100098, 
and
No.~LE230100085; 
Austrian Federal Ministry of Education, Science and Research,
Austrian Science Fund
No.~P~34529,
No.~J~4731,
No.~J~4625,
and
No.~M~3153,
and
Horizon 2020 ERC Starting Grant No.~947006 ``InterLeptons'';
Natural Sciences and Engineering Research Council of Canada, Compute Canada and CANARIE;
National Key R\&D Program of China under Contract No.~2022YFA1601903,
National Natural Science Foundation of China and Research Grants
No.~11575017,
No.~11761141009,
No.~11705209,
No.~11975076,
No.~12135005,
No.~12150004,
No.~12161141008,
and
No.~12175041,
and Shandong Provincial Natural Science Foundation Project~ZR2022JQ02;
the Czech Science Foundation Grant No.~22-18469S 
and
Charles University Grant Agency project No.~246122;
European Research Council, Seventh Framework PIEF-GA-2013-622527,
Horizon 2020 ERC-Advanced Grants No.~267104 and No.~884719,
Horizon 2020 ERC-Consolidator Grant No.~819127,
Horizon 2020 Marie Sklodowska-Curie Grant Agreement No.~700525 ``NIOBE''
and
No.~101026516,
and
Horizon 2020 Marie Sklodowska-Curie RISE project JENNIFER2 Grant Agreement No.~822070 (European grants);
L'Institut National de Physique Nucl\'{e}aire et de Physique des Particules (IN2P3) du CNRS
and
L'Agence Nationale de la Recherche (ANR) under grant ANR-21-CE31-0009 (France);
BMBF, DFG, HGF, MPG, and AvH Foundation (Germany);
Department of Atomic Energy under Project Identification No.~RTI 4002,
Department of Science and Technology,
and
UPES SEED funding programs
No.~UPES/R\&D-SEED-INFRA/17052023/01 and
No.~UPES/R\&D-SOE/20062022/06 (India);
Israel Science Foundation Grant No.~2476/17,
U.S.-Israel Binational Science Foundation Grant No.~2016113, and
Israel Ministry of Science Grant No.~3-16543;
Istituto Nazionale di Fisica Nucleare and the Research Grants BELLE2;
Japan Society for the Promotion of Science, Grant-in-Aid for Scientific Research Grants
No.~16H03968,
No.~16H03993,
No.~16H06492,
No.~16K05323,
No.~17H01133,
No.~17H05405,
No.~18K03621,
No.~18H03710,
No.~18H05226,
No.~19H00682, 
No.~20H05850,
No.~20H05858,
No.~22H00144,
No.~22K14056,
No.~22K21347,
No.~23H05433,
No.~26220706,
and
No.~26400255,
and
the Ministry of Education, Culture, Sports, Science, and Technology (MEXT) of Japan;  
National Research Foundation (NRF) of Korea Grants
No.~2016R1\-D1A1B\-02012900,
No.~2018R1\-A2B\-3003643,
No.~2018R1\-A6A1A\-06024970,
No.~2019R1\-I1A3A\-01058933,
No.~2021R1\-A6A1A\-03043957,
No.~2021R1\-F1A\-1060423,
No.~2021R1\-F1A\-1064008,
No.~2022R1\-A2C\-1003993,
and
No.~RS-2022-00197659,
Radiation Science Research Institute,
Foreign Large-Size Research Facility Application Supporting project,
the Global Science Experimental Data Hub Center of the Korea Institute of Science and Technology Information
and
KREONET/GLORIAD;
Universiti Malaya RU grant, Akademi Sains Malaysia, and Ministry of Education Malaysia;
Frontiers of Science Program Contracts
No.~FOINS-296,
No.~CB-221329,
No.~CB-236394,
No.~CB-254409,
and
No.~CB-180023, and SEP-CINVESTAV Research Grant No.~237 (Mexico);
the Polish Ministry of Science and Higher Education and the National Science Center;
the Ministry of Science and Higher Education of the Russian Federation
and
the HSE University Basic Research Program, Moscow;
University of Tabuk Research Grants
No.~S-0256-1438 and No.~S-0280-1439 (Saudi Arabia);
Slovenian Research Agency and Research Grants
No.~J1-9124
and
No.~P1-0135;
Agencia Estatal de Investigacion, Spain
Grant No.~RYC2020-029875-I
and
Generalitat Valenciana, Spain
Grant No.~CIDEGENT/2018/020;
National Science and Technology Council,
and
Ministry of Education (Taiwan);
Thailand Center of Excellence in Physics;
TUBITAK ULAKBIM (Turkey);
National Research Foundation of Ukraine, Project No.~2020.02/0257,
and
Ministry of Education and Science of Ukraine;
the U.S. National Science Foundation and Research Grants
No.~PHY-1913789 
and
No.~PHY-2111604, 
and the U.S. Department of Energy and Research Awards
No.~DE-AC06-76RLO1830, 
No.~DE-SC0007983, 
No.~DE-SC0009824, 
No.~DE-SC0009973, 
No.~DE-SC0010007, 
No.~DE-SC0010073, 
No.~DE-SC0010118, 
No.~DE-SC0010504, 
No.~DE-SC0011784, 
No.~DE-SC0012704, 
No.~DE-SC0019230, 
No.~DE-SC0021274, 
No.~DE-SC0021616, 
No.~DE-SC0022350, 
No.~DE-SC0023470; 
and
the Vietnam Academy of Science and Technology (VAST) under Grants
No.~NVCC.05.12/22-23
and
No.~DL0000.02/24-25.

These acknowledgements are not to be interpreted as an endorsement of any statement made
by any of our institutes, funding agencies, governments, or their representatives.

We thank the SuperKEKB team for delivering high-luminosity collisions;
the KEK cryogenics group for the efficient operation of the detector solenoid magnet;
the KEK Computer Research Center for on-site computing support; the NII for SINET6 network support;
and the raw-data centers hosted by BNL, DESY, GridKa, IN2P3, INFN, 
and the University of Victoria.

\bibliography{references}

\begin{thebibliography}{51}%
\makeatletter
\providecommand \@ifxundefined [1]{%
 \@ifx{#1\undefined}
}%
\providecommand \@ifnum [1]{%
 \ifnum #1\expandafter \@firstoftwo
 \else \expandafter \@secondoftwo
 \fi
}%
\providecommand \@ifx [1]{%
 \ifx #1\expandafter \@firstoftwo
 \else \expandafter \@secondoftwo
 \fi
}%
\providecommand \natexlab [1]{#1}%
\providecommand \enquote  [1]{``#1''}%
\providecommand \bibnamefont  [1]{#1}%
\providecommand \bibfnamefont [1]{#1}%
\providecommand \citenamefont [1]{#1}%
\providecommand \href@noop [0]{\@secondoftwo}%
\providecommand \href [0]{\begingroup \@sanitize@url \@href}%
\providecommand \@href[1]{\@@startlink{#1}\@@href}%
\providecommand \@@href[1]{\endgroup#1\@@endlink}%
\providecommand \@sanitize@url [0]{\catcode `\\12\catcode `\$12\catcode
  `\&12\catcode `\#12\catcode `\^12\catcode `\_12\catcode `\%12\relax}%
\providecommand \@@startlink[1]{}%
\providecommand \@@endlink[0]{}%
\providecommand \url  [0]{\begingroup\@sanitize@url \@url }%
\providecommand \@url [1]{\endgroup\@href {#1}{\urlprefix }}%
\providecommand \urlprefix  [0]{URL }%
\providecommand \Eprint [0]{\href }%
\providecommand \doibase [0]{https://doi.org/}%
\providecommand \selectlanguage [0]{\@gobble}%
\providecommand \bibinfo  [0]{\@secondoftwo}%
\providecommand \bibfield  [0]{\@secondoftwo}%
\providecommand \translation [1]{[#1]}%
\providecommand \BibitemOpen [0]{}%
\providecommand \bibitemStop [0]{}%
\providecommand \bibitemNoStop [0]{.\EOS\space}%
\providecommand \EOS [0]{\spacefactor3000\relax}%
\providecommand \BibitemShut  [1]{\csname bibitem#1\endcsname}%
\let\auto@bib@innerbib\@empty
\bibitem [{\citenamefont {Bernlochner}\ \emph
  {et~al.}(2022{\natexlab{a}})\citenamefont {Bernlochner} \emph
  {et~al.}}]{bernlochner_review}%
  \BibitemOpen
  \bibfield  {author} {\bibinfo {author} {\bibfnamefont {F.~U.}\ \bibnamefont
  {Bernlochner}} \emph {et~al.},\ }\bibfield  {title} {\bibinfo {title}
  {{Semitauonic $b$-hadron decays: A lepton flavor universality laboratory}},\
  }\href {https://doi.org/10.1103/RevModPhys.94.015003} {\bibfield  {journal}
  {\bibinfo  {journal} {Rev. Mod. Phys.}\ }\textbf {\bibinfo {volume} {94}},\
  \bibinfo {pages} {015003} (\bibinfo {year} {2022}{\natexlab{a}})}\BibitemShut
  {NoStop}%
\bibitem [{\citenamefont {Lees}\ \emph {et~al.}(2012)\citenamefont {Lees} \emph
  {et~al.}}]{babar_1}%
  \BibitemOpen
  \bibfield  {author} {\bibinfo {author} {\bibfnamefont {J.~P.}\ \bibnamefont
  {Lees}} \emph {et~al.} (\bibinfo {collaboration} {BaBar Collaboration}),\
  }\bibfield  {title} {\bibinfo {title} {{{Evidence for an excess of $\Bbar \to
  D^{(*)} \tau^-\bar{\nu}_\tau$ decays}}},\ }\href
  {https://doi.org/10.1103/PhysRevLett.109.101802} {\bibfield  {journal}
  {\bibinfo  {journal} {Phys. Rev. Lett.}\ }\textbf {\bibinfo {volume} {109}},\
  \bibinfo {pages} {101802} (\bibinfo {year} {2012})}\BibitemShut {NoStop}%
\bibitem [{\citenamefont {Lees}\ \emph {et~al.}(2013)\citenamefont {Lees} \emph
  {et~al.}}]{babar_2}%
  \BibitemOpen
  \bibfield  {author} {\bibinfo {author} {\bibfnamefont {J.~P.}\ \bibnamefont
  {Lees}} \emph {et~al.} (\bibinfo {collaboration} {BaBar Collaboration}),\
  }\bibfield  {title} {\bibinfo {title} {{{Measurement of an excess of $\Bbar
  \to D^{(*)}\tau^- \bar{\nu}_\tau$ decays and implications for charged Higgs
  bosons}}},\ }\href {https://doi.org/10.1103/PhysRevD.88.072012} {\bibfield
  {journal} {\bibinfo  {journal} {Phys. Rev. D}\ }\textbf {\bibinfo {volume}
  {88}},\ \bibinfo {pages} {072012} (\bibinfo {year} {2013})}\BibitemShut
  {NoStop}%
\bibitem [{\citenamefont {Huschle}\ \emph {et~al.}(2015)\citenamefont {Huschle}
  \emph {et~al.}}]{belle_hadronic}%
  \BibitemOpen
  \bibfield  {author} {\bibinfo {author} {\bibfnamefont {M.}~\bibnamefont
  {Huschle}} \emph {et~al.} (\bibinfo {collaboration} {Belle Collaboration}),\
  }\bibfield  {title} {\bibinfo {title} {{{Measurement of the branching ratio
  of $\Bbar \to D^{(\ast)} \tau^- \bar{\nu}_\tau$ relative to $\Bbar \to
  D^{(\ast)} \ell^- \bar{\nu}_\ell$ decays with hadronic tagging at Belle}}},\
  }\href {https://doi.org/10.1103/PhysRevD.92.072014} {\bibfield  {journal}
  {\bibinfo  {journal} {Phys. Rev. D}\ }\textbf {\bibinfo {volume} {92}},\
  \bibinfo {pages} {072014} (\bibinfo {year} {2015})}\BibitemShut {NoStop}%
\bibitem [{\citenamefont {Hirose}\ \emph {et~al.}(2017)\citenamefont {Hirose}
  \emph {et~al.}}]{belle_pol_PRL}%
  \BibitemOpen
  \bibfield  {author} {\bibinfo {author} {\bibfnamefont {S.}~\bibnamefont
  {Hirose}} \emph {et~al.} (\bibinfo {collaboration} {Belle Collaboration}),\
  }\bibfield  {title} {\bibinfo {title} {{{Measurement of the $\tau$ lepton
  polarization and $R(D^*)$ in the decay $\Bbar \to D^* \tau^-
  \bar{\nu}_\tau$}}},\ }\href {https://doi.org/10.1103/PhysRevLett.118.211801}
  {\bibfield  {journal} {\bibinfo  {journal} {Phys. Rev. Lett.}\ }\textbf
  {\bibinfo {volume} {118}},\ \bibinfo {pages} {211801} (\bibinfo {year}
  {2017})}\BibitemShut {NoStop}%
\bibitem [{\citenamefont {Hirose}\ \emph {et~al.}(2018)\citenamefont {Hirose}
  \emph {et~al.}}]{belle_pol_PRD}%
  \BibitemOpen
  \bibfield  {author} {\bibinfo {author} {\bibfnamefont {S.}~\bibnamefont
  {Hirose}} \emph {et~al.} (\bibinfo {collaboration} {Belle Collaboration}),\
  }\bibfield  {title} {\bibinfo {title} {{Measurement of the $\tau$ lepton
  polarization and $R(D^*)$ in the decay $\Bbar\to D^* \tau^- \bar{\nu}_\tau$
  with one-prong hadronic $\tau$ decays at Belle}},\ }\href
  {https://doi.org/10.1103/PhysRevD.97.012004} {\bibfield  {journal} {\bibinfo
  {journal} {Phys. Rev. D}\ }\textbf {\bibinfo {volume} {97}},\ \bibinfo
  {pages} {012004} (\bibinfo {year} {2018})}\BibitemShut {NoStop}%
\bibitem [{\citenamefont {Caria}\ \emph {et~al.}(2020)\citenamefont {Caria}
  \emph {et~al.}}]{belle_semileptonic}%
  \BibitemOpen
  \bibfield  {author} {\bibinfo {author} {\bibfnamefont {G.}~\bibnamefont
  {Caria}} \emph {et~al.} (\bibinfo {collaboration} {Belle Collaboration}),\
  }\bibfield  {title} {\bibinfo {title} {{Measurement of $R(D)$ and
  $R({D}^{*})$ with a semileptonic tagging method}},\ }\href
  {https://doi.org/10.1103/PhysRevLett.124.161803} {\bibfield  {journal}
  {\bibinfo  {journal} {Phys. Rev. Lett.}\ }\textbf {\bibinfo {volume} {124}},\
  \bibinfo {pages} {161803} (\bibinfo {year} {2020})}\BibitemShut {NoStop}%
\bibitem [{\citenamefont {Adachi}\ \emph {et~al.}(2024)\citenamefont {Adachi}
  \emph {et~al.}}]{b2_rdst}%
  \BibitemOpen
  \bibfield  {author} {\bibinfo {author} {\bibfnamefont {I.}~\bibnamefont
  {Adachi}} \emph {et~al.} (\bibinfo {collaboration} {Belle II
  Collaboration}),\ }\bibfield  {title} {\bibinfo {title} {{A test of lepton
  flavor universality with a measurement of $R(D^{*})$ using hadronic $B$
  tagging at the Belle II experiment}},\ }\href@noop {} {\  (\bibinfo {year}
  {2024})},\ \Eprint {https://arxiv.org/abs/2401.02840} {arXiv:2401.02840
  [hep-ex]} \BibitemShut {NoStop}%
\bibitem [{\citenamefont {Aaij}\ \emph
  {et~al.}(2023{\natexlab{a}})\citenamefont {Aaij} \emph {et~al.}}]{lhcb_new1}%
  \BibitemOpen
  \bibfield  {author} {\bibinfo {author} {\bibfnamefont {R.}~\bibnamefont
  {Aaij}} \emph {et~al.} (\bibinfo {collaboration} {LHCb Collaboration}),\
  }\bibfield  {title} {\bibinfo {title} {{Test of lepton flavor universality
  using $B^0 \to D^{*-}\tau^+\nu_{\tau}$ decays with hadronic $\tau$
  channels}},\ }\href {https://doi.org/10.1103/PhysRevD.108.012018} {\bibfield
  {journal} {\bibinfo  {journal} {Phys. Rev. D}\ }\textbf {\bibinfo {volume}
  {108}},\ \bibinfo {pages} {012018} (\bibinfo {year}
  {2023}{\natexlab{a}})}\BibitemShut {NoStop}%
\bibitem [{\citenamefont {Aaij}\ \emph
  {et~al.}(2023{\natexlab{b}})\citenamefont {Aaij} \emph {et~al.}}]{lhcb_new2}%
  \BibitemOpen
  \bibfield  {author} {\bibinfo {author} {\bibfnamefont {R.}~\bibnamefont
  {Aaij}} \emph {et~al.} (\bibinfo {collaboration} {LHCb Collaboration}),\
  }\bibfield  {title} {\bibinfo {title} {{Measurement of the ratios of
  branching fractions $\mathcal{R}(D^{*})$ and $\mathcal{R}(D^{0})$}},\ }\href
  {https://doi.org/10.1103/PhysRevLett.131.111802} {\bibfield  {journal}
  {\bibinfo  {journal} {Phys. Rev. Lett.}\ }\textbf {\bibinfo {volume} {131}},\
  \bibinfo {pages} {111802} (\bibinfo {year} {2023}{\natexlab{b}})}\BibitemShut
  {NoStop}%
\bibitem [{\citenamefont {{Heavy Flavor Averaging Group
  Collaboration}}(2023)}]{hflav2023}%
  \BibitemOpen
  \bibfield  {author} {\bibinfo {author} {\bibnamefont {{Heavy Flavor Averaging
  Group Collaboration}}},\ }\href
  {https://hflav-eos.web.cern.ch/hflav-eos/semi/summer23/html/RDsDsstar/RDRDs.html}
  {\bibinfo {title} {{Preliminary average of $R(D)$ and $R(D^*)$ for Summer
  2023}}} (\bibinfo {year} {2023}),\ \bibinfo {note} {visited on
  12/10/2023}\BibitemShut {NoStop}%
\bibitem [{\citenamefont {Grossman}\ and\ \citenamefont
  {Ligeti}(1994)}]{twoHDM}%
  \BibitemOpen
  \bibfield  {author} {\bibinfo {author} {\bibfnamefont {Y.}~\bibnamefont
  {Grossman}}\ and\ \bibinfo {author} {\bibfnamefont {Z.}~\bibnamefont
  {Ligeti}},\ }\bibfield  {title} {\bibinfo {title} {{{The inclusive $B
  \to\tau\nu X$ decay in two Higgs doublet models}}},\ }\href
  {https://doi.org/10.1016/0370-2693(94)91267-X} {\bibfield  {journal}
  {\bibinfo  {journal} {Phys. Lett. B}\ }\textbf {\bibinfo {volume} {332}},\
  \bibinfo {pages} {373} (\bibinfo {year} {1994})}\BibitemShut {NoStop}%
\bibitem [{\citenamefont {Be\ifmmode \check{c}\else
  \v{c}\fi{}irevi\ifmmode~\acute{c}\else \'{c}\fi{}}\ \emph
  {et~al.}(2018)\citenamefont {Be\ifmmode \check{c}\else
  \v{c}\fi{}irevi\ifmmode~\acute{c}\else \'{c}\fi{}} \emph {et~al.}}]{SLQ}%
  \BibitemOpen
  \bibfield  {author} {\bibinfo {author} {\bibfnamefont {D.}~\bibnamefont
  {Be\ifmmode \check{c}\else \v{c}\fi{}irevi\ifmmode~\acute{c}\else
  \'{c}\fi{}}} \emph {et~al.},\ }\bibfield  {title} {\bibinfo {title} {{Scalar
  leptoquarks from grand unified theories to accommodate the $B$-physics
  anomalies}},\ }\href {https://doi.org/10.1103/PhysRevD.98.055003} {\bibfield
  {journal} {\bibinfo  {journal} {Phys. Rev. D}\ }\textbf {\bibinfo {volume}
  {98}},\ \bibinfo {pages} {055003} (\bibinfo {year} {2018})}\BibitemShut
  {NoStop}%
\bibitem [{\citenamefont {Barate}\ \emph {et~al.}(2001)\citenamefont {Barate}
  \emph {et~al.}}]{aleph}%
  \BibitemOpen
  \bibfield  {author} {\bibinfo {author} {\bibfnamefont {R.}~\bibnamefont
  {Barate}} \emph {et~al.} (\bibinfo {collaboration} {ALEPH Collaboration}),\
  }\bibfield  {title} {\bibinfo {title} {{{Measurements of $\mathcal{B}(b
  \rightarrow \tau^- \bar{\nu}_\tau X)$ and $\mathcal{B}(b \rightarrow \tau^-
  \bar{\nu}_\tau D^{*\pm} X)$ and upper limits on $\mathcal{B} (B^- \rightarrow
  \tau^- \bar{\nu}_\tau)$ and $\mathcal{B} (b\rightarrow s \nu \bar{\nu})$}}},\
  }\href {https://doi.org/10.1007/s100520100612} {\bibfield  {journal}
  {\bibinfo  {journal} {Eur. Phys. J. C}\ }\textbf {\bibinfo {volume} {19}},\
  \bibinfo {pages} {213} (\bibinfo {year} {2001})}\BibitemShut {NoStop}%
\bibitem [{\citenamefont {Abreu}\ \emph {et~al.}(2000)\citenamefont {Abreu}
  \emph {et~al.}}]{delphi}%
  \BibitemOpen
  \bibfield  {author} {\bibinfo {author} {\bibfnamefont {P.}~\bibnamefont
  {Abreu}} \emph {et~al.} (\bibinfo {collaboration} {DELPHI Collaboration}),\
  }\bibfield  {title} {\bibinfo {title} {{{Upper limit for the decay $B^{-} \to
  \tau^{-} \bar{\nu}_\tau$ and measurement of the $b \to \tau \bar{\nu}_\tau X$
  branching ratio}}},\ }\href {https://doi.org/10.1016/S0370-2693(00)01274-0}
  {\bibfield  {journal} {\bibinfo  {journal} {Phys. Lett. B}\ }\textbf
  {\bibinfo {volume} {496}},\ \bibinfo {pages} {43} (\bibinfo {year}
  {2000})}\BibitemShut {NoStop}%
\bibitem [{\citenamefont {Acciarri}\ \emph {et~al.}(1994)\citenamefont
  {Acciarri} \emph {et~al.}}]{l3_1}%
  \BibitemOpen
  \bibfield  {author} {\bibinfo {author} {\bibfnamefont {M.}~\bibnamefont
  {Acciarri}} \emph {et~al.} (\bibinfo {collaboration} {L3 Collaboration}),\
  }\bibfield  {title} {\bibinfo {title} {{{Measurement of the inclusive $B
  \rightarrow \tau\nu X$ branching ratio}}},\ }\href
  {https://doi.org/10.1016/0370-2693(94)90880-X} {\bibfield  {journal}
  {\bibinfo  {journal} {Phys. Lett. B}\ }\textbf {\bibinfo {volume} {332}},\
  \bibinfo {pages} {201} (\bibinfo {year} {1994})}\BibitemShut {NoStop}%
\bibitem [{\citenamefont {Acciarri}\ \emph {et~al.}(1996)\citenamefont
  {Acciarri} \emph {et~al.}}]{l3_2}%
  \BibitemOpen
  \bibfield  {author} {\bibinfo {author} {\bibfnamefont {M.}~\bibnamefont
  {Acciarri}} \emph {et~al.} (\bibinfo {collaboration} {L3 Collaboration}),\
  }\bibfield  {title} {\bibinfo {title} {{{Measurement of the branching ratios
  $b \rightarrow e \nu X, \mu \nu X, \tau\nu X$ and $\nu X$}}},\ }\href
  {http://cds.cern.ch/record/302713} {\bibfield  {journal} {\bibinfo  {journal}
  {Z. Phys. C}\ }\textbf {\bibinfo {volume} {71}},\ \bibinfo {pages} {379}
  (\bibinfo {year} {1996})}\BibitemShut {NoStop}%
\bibitem [{\citenamefont {Abbiendi}\ \emph {et~al.}(2001)\citenamefont
  {Abbiendi} \emph {et~al.}}]{opal}%
  \BibitemOpen
  \bibfield  {author} {\bibinfo {author} {\bibfnamefont {G.}~\bibnamefont
  {Abbiendi}} \emph {et~al.} (\bibinfo {collaboration} {OPAL Collaboration}),\
  }\bibfield  {title} {\bibinfo {title} {{{Measurement of the branching ratio
  for the process $b\rightarrow \tau^- \bar{\nu}_\tau X$}}},\ }\href
  {https://doi.org/10.1016/S0370-2693(01)01012-7} {\bibfield  {journal}
  {\bibinfo  {journal} {Phys. Lett. B}\ }\textbf {\bibinfo {volume} {520}},\
  \bibinfo {pages} {1} (\bibinfo {year} {2001})}\BibitemShut {NoStop}%
\bibitem [{sup()}]{supplemental}%
  \BibitemOpen
  \href@noop {} {\bibinfo  {journal} {See Supplemental Material below for
  additional Figures validating the simulation reweighting and information on
  the constraints on \RDorDst implied by \RX, which includes Ref.
  \cite{mannel_rxc}}\ }\BibitemShut {NoStop}%
\bibitem [{\citenamefont {Mannel}\ \emph {et~al.}(2017)\citenamefont {Mannel},
  \citenamefont {Rusov},\ and\ \citenamefont {Shahriaran}}]{mannel_rxc}%
  \BibitemOpen
\bibfield  {journal} {  }\bibfield  {author} {\bibinfo {author} {\bibfnamefont
  {T.}~\bibnamefont {Mannel}}, \bibinfo {author} {\bibfnamefont {A.~V.}\
  \bibnamefont {Rusov}},\ and\ \bibinfo {author} {\bibfnamefont
  {F.}~\bibnamefont {Shahriaran}},\ }\bibfield  {title} {\bibinfo {title}
  {{Inclusive semitauonic $B$ decays to order
  $\mathcal{O}(\Lambda_\mathrm{QCD}^3/m_b^3)$}},\ }\href
  {https://doi.org/10.1016/j.nuclphysb.2017.05.016} {\bibfield  {journal}
  {\bibinfo  {journal} {Nucl. Phys. B}\ }\textbf {\bibinfo {volume} {921}},\
  \bibinfo {pages} {211} (\bibinfo {year} {2017})}\BibitemShut {NoStop}%
\bibitem [{\citenamefont {Rahimi}\ and\ \citenamefont {Vos}(2022)}]{kvos}%
  \BibitemOpen
  \bibfield  {author} {\bibinfo {author} {\bibfnamefont {M.}~\bibnamefont
  {Rahimi}}\ and\ \bibinfo {author} {\bibfnamefont {K.~K.}\ \bibnamefont
  {Vos}},\ }\bibfield  {title} {\bibinfo {title} {{Standard model predictions
  for lepton flavour universality ratios of inclusive semileptonic $B$
  decays}},\ }\href {https://doi.org/10.1007/JHEP11(2022)007} {\bibfield
  {journal} {\bibinfo  {journal} {J. High Energ. Phys.}\ }\textbf {\bibinfo
  {volume} {2022}},\ \bibinfo {pages} {7 (2022)}}\BibitemShut {NoStop}%
\bibitem [{\citenamefont {Freytsis}\ \emph {et~al.}(2015)\citenamefont
  {Freytsis}, \citenamefont {Ligeti},\ and\ \citenamefont
  {Ruderman}}]{ligeti_rxc}%
  \BibitemOpen
  \bibfield  {author} {\bibinfo {author} {\bibfnamefont {M.}~\bibnamefont
  {Freytsis}}, \bibinfo {author} {\bibfnamefont {Z.}~\bibnamefont {Ligeti}},\
  and\ \bibinfo {author} {\bibfnamefont {J.~T.}\ \bibnamefont {Ruderman}},\
  }\bibfield  {title} {\bibinfo {title} {{Flavor models for $\Bbar
  \ensuremath{\rightarrow}{D}^{(*)}\ensuremath{\tau}\overline{\ensuremath{\nu}}$}},\
  }\href {https://doi.org/10.1103/PhysRevD.92.054018} {\bibfield  {journal}
  {\bibinfo  {journal} {Phys. Rev. D}\ }\textbf {\bibinfo {volume} {92}},\
  \bibinfo {pages} {054018} (\bibinfo {year} {2015})}\BibitemShut {NoStop}%
\bibitem [{\citenamefont {Aggarwal}\ \emph {et~al.}(2022)\citenamefont
  {Aggarwal} \emph {et~al.}}]{snowmass}%
  \BibitemOpen
  \bibfield  {author} {\bibinfo {author} {\bibfnamefont {L.}~\bibnamefont
  {Aggarwal}} \emph {et~al.},\ }\bibfield  {title} {\bibinfo {title} {{Snowmass
  white paper: Belle II physics reach and plans for the next decade and
  beyond}},\ }\href@noop {} {\  (\bibinfo {year} {2022})},\ \Eprint
  {https://arxiv.org/abs/2207.06307} {arXiv:2207.06307 [hep-ex]} \BibitemShut
  {NoStop}%
\bibitem [{\citenamefont {Akai}\ \emph {et~al.}(2018)\citenamefont {Akai} \emph
  {et~al.}}]{superkekb}%
  \BibitemOpen
  \bibfield  {author} {\bibinfo {author} {\bibfnamefont {K.}~\bibnamefont
  {Akai}} \emph {et~al.} (\bibinfo {collaboration} {SuperKEKB Accelerator
  Team}),\ }\bibfield  {title} {\bibinfo {title} {{{SuperKEKB collider}}},\
  }\href {https://doi.org/10.1016/j.nima.2018.08.017} {\bibfield  {journal}
  {\bibinfo  {journal} {Nucl. Instrum. Meth. A}\ }\textbf {\bibinfo {volume}
  {907}},\ \bibinfo {pages} {188} (\bibinfo {year} {2018})}\BibitemShut
  {NoStop}%
\bibitem [{\citenamefont {Abe}\ \emph {et~al.}(2010)\citenamefont {Abe} \emph
  {et~al.}}]{b2tdr}%
  \BibitemOpen
  \bibfield  {author} {\bibinfo {author} {\bibfnamefont {T.}~\bibnamefont
  {Abe}} \emph {et~al.} (\bibinfo {collaboration} {Belle II Collaboration}),\
  }\bibfield  {title} {\bibinfo {title} {{{Belle II technical design
  report}}},\ }\href@noop {} {\  (\bibinfo {year} {2010})},\ \Eprint
  {https://arxiv.org/abs/1011.0352} {arXiv:1011.0352 [physics.ins-det]}
  \BibitemShut {NoStop}%
\bibitem [{\citenamefont {Altmannshofer}\ \emph {et~al.}(2019)\citenamefont
  {Altmannshofer} \emph {et~al.}}]{b2tip}%
  \BibitemOpen
  \bibfield  {author} {\bibinfo {author} {\bibfnamefont {W.}~\bibnamefont
  {Altmannshofer}} \emph {et~al.},\ }\bibfield  {title} {\bibinfo {title} {{The
  Belle II physics book}},\ }\href {https://doi.org/10.1093/ptep/ptz106}
  {\bibfield  {journal} {\bibinfo  {journal} {Prog. Theor. Exp. Phys.}\
  }\textbf {\bibinfo {volume} {2019}},\ \bibinfo {pages} {123C01} (\bibinfo
  {year} {2019})},\ \bibinfo {note} {[Erratum: Prog. Theor. Exp. Phys.
  \textbf{2020}, 029201 (2020)]}\BibitemShut {NoStop}%
\bibitem [{\citenamefont {Lange}(2001)}]{evtgen}%
  \BibitemOpen
  \bibfield  {author} {\bibinfo {author} {\bibfnamefont {D.~J.}\ \bibnamefont
  {Lange}},\ }\bibfield  {title} {\bibinfo {title} {{The EvtGen particle decay
  simulation package}},\ }\href {https://doi.org/10.1016/S0168-9002(01)00089-4}
  {\bibfield  {journal} {\bibinfo  {journal} {Nucl. Instrum. Meth. A}\ }\textbf
  {\bibinfo {volume} {462}},\ \bibinfo {pages} {152} (\bibinfo {year}
  {2001})}\BibitemShut {NoStop}%
\bibitem [{\citenamefont {Sjöstrand}\ \emph {et~al.}(2015)\citenamefont
  {Sjöstrand} \emph {et~al.}}]{pythia8}%
  \BibitemOpen
  \bibfield  {author} {\bibinfo {author} {\bibfnamefont {T.}~\bibnamefont
  {Sjöstrand}} \emph {et~al.},\ }\bibfield  {title} {\bibinfo {title} {{An
  introduction to PYTHIA 8.2}},\ }\href
  {https://doi.org/10.1016/j.cpc.2015.01.024} {\bibfield  {journal} {\bibinfo
  {journal} {Comput. Phys. Commun.}\ }\textbf {\bibinfo {volume} {191}},\
  \bibinfo {pages} {159} (\bibinfo {year} {2015})}\BibitemShut {NoStop}%
\bibitem [{\citenamefont {Jadach}\ \emph {et~al.}(2000)\citenamefont {Jadach}
  \emph {et~al.}}]{kkmc}%
  \BibitemOpen
  \bibfield  {author} {\bibinfo {author} {\bibfnamefont {S.}~\bibnamefont
  {Jadach}} \emph {et~al.},\ }\bibfield  {title} {\bibinfo {title} {{{The
  precision Monte Carlo event generator KK for two fermion final states in $e^+
  e^-$ collisions}}},\ }\href {https://doi.org/10.1016/S0010-4655(00)00048-5}
  {\bibfield  {journal} {\bibinfo  {journal} {Comput. Phys. Commun.}\ }\textbf
  {\bibinfo {volume} {130}},\ \bibinfo {pages} {260} (\bibinfo {year}
  {2000})}\BibitemShut {NoStop}%
\bibitem [{\citenamefont {Barberio}\ \emph {et~al.}(1991)\citenamefont
  {Barberio}, \citenamefont {{van Eijk}},\ and\ \citenamefont {Was}}]{PHOTOS}%
  \BibitemOpen
  \bibfield  {author} {\bibinfo {author} {\bibfnamefont {E.}~\bibnamefont
  {Barberio}}, \bibinfo {author} {\bibfnamefont {B.}~\bibnamefont {{van
  Eijk}}},\ and\ \bibinfo {author} {\bibfnamefont {Z.}~\bibnamefont {Was}},\
  }\bibfield  {title} {\bibinfo {title} {{Photos - a universal Monte Carlo for
  QED radiative corrections in decays}},\ }\href
  {https://doi.org/10.1016/0010-4655(91)90012-A} {\bibfield  {journal}
  {\bibinfo  {journal} {Comput. Phys. Commun.}\ }\textbf {\bibinfo {volume}
  {66}},\ \bibinfo {pages} {115} (\bibinfo {year} {1991})}\BibitemShut
  {NoStop}%
\bibitem [{\citenamefont {Agostinelli}\ \emph {et~al.}(2003)\citenamefont
  {Agostinelli} \emph {et~al.}}]{geant4}%
  \BibitemOpen
  \bibfield  {author} {\bibinfo {author} {\bibfnamefont {S.}~\bibnamefont
  {Agostinelli}} \emph {et~al.} (\bibinfo {collaboration} {\textsc{Geant4}
  Collaboration}),\ }\bibfield  {title} {\bibinfo {title}
  {{{\textsc{Geant4}—a simulation toolkit}}},\ }\href
  {https://doi.org/10.1016/S0168-9002(03)01368-8} {\bibfield  {journal}
  {\bibinfo  {journal} {Nucl. Instrum. Meth. A}\ }\textbf {\bibinfo {volume}
  {506}},\ \bibinfo {pages} {250} (\bibinfo {year} {2003})}\BibitemShut
  {NoStop}%
\bibitem [{\citenamefont {Natochii}\ \emph {et~al.}(2022)\citenamefont
  {Natochii} \emph {et~al.}}]{BeamBKG}%
  \BibitemOpen
  \bibfield  {author} {\bibinfo {author} {\bibfnamefont {A.}~\bibnamefont
  {Natochii}} \emph {et~al.},\ }\bibfield  {title} {\bibinfo {title} {{Beam
  background expectations for {Belle II at SuperKEKB}}},\ }\href@noop {} {\
  (\bibinfo {year} {2022})},\ \Eprint {https://arxiv.org/abs/2203.05731}
  {arXiv:2203.05731 [hep-ex]} \BibitemShut {NoStop}%
\bibitem [{\citenamefont {Kuhr}\ \emph {et~al.}(2019)\citenamefont {Kuhr} \emph
  {et~al.}}]{basf2}%
  \BibitemOpen
  \bibfield  {author} {\bibinfo {author} {\bibfnamefont {T.}~\bibnamefont
  {Kuhr}} \emph {et~al.} (\bibinfo {collaboration} {Belle II Framework Software
  Group}),\ }\bibfield  {title} {\bibinfo {title} {{{The Belle II core
  software}}},\ }\href {https://doi.org/10.1007/s41781-018-0017-9} {\bibfield
  {journal} {\bibinfo  {journal} {Comput. Softw. Big Sci.}\ }\textbf {\bibinfo
  {volume} {3}},\ \bibinfo {pages} {1} (\bibinfo {year} {2019})}\BibitemShut
  {NoStop}%
\bibitem [{\citenamefont {{Belle II Collaboration}}()}]{basf2-zenodo}%
  \BibitemOpen
  \bibfield  {author} {\bibinfo {author} {\bibnamefont {{Belle II
  Collaboration}}},\ }\bibfield  {title} {\bibinfo {title} {{Belle II Analysis
  Software Framework (basf2)}}\ }\href {https://doi.org/10.5281/zenodo.5574115}
  {10.5281/zenodo.5574115}\BibitemShut {NoStop}%
\bibitem [{\citenamefont {Workman}\ \emph {et~al.}(2022)\citenamefont {Workman}
  \emph {et~al.}}]{pdg}%
  \BibitemOpen
  \bibfield  {author} {\bibinfo {author} {\bibfnamefont {R.~L.}\ \bibnamefont
  {Workman}} \emph {et~al.} (\bibinfo {collaboration} {Particle Data Group}),\
  }\bibfield  {title} {\bibinfo {title} {{Review of particle physics}},\ }\href
  {https://doi.org/10.1093/ptep/ptac097} {\bibfield  {journal} {\bibinfo
  {journal} {Prog. Theor. Exp. Phys.}\ }\textbf {\bibinfo {volume} {2022}},\
  \bibinfo {pages} {083C01} (\bibinfo {year} {2022})}\BibitemShut {NoStop}%
\bibitem [{\citenamefont {Bernlochner}\ \emph
  {et~al.}(2022{\natexlab{b}})\citenamefont {Bernlochner} \emph
  {et~al.}}]{blprxp}%
  \BibitemOpen
  \bibfield  {author} {\bibinfo {author} {\bibfnamefont {F.~U.}\ \bibnamefont
  {Bernlochner}} \emph {et~al.},\ }\bibfield  {title} {\bibinfo {title}
  {{Constrained second-order power corrections in HQET: $R({D}^{(*)})$,
  $|{V}_{cb}|$, and new physics}},\ }\href
  {https://doi.org/10.1103/PhysRevD.106.096015} {\bibfield  {journal} {\bibinfo
   {journal} {Phys. Rev. D}\ }\textbf {\bibinfo {volume} {106}},\ \bibinfo
  {pages} {096015} (\bibinfo {year} {2022}{\natexlab{b}})}\BibitemShut
  {NoStop}%
\bibitem [{\citenamefont {Bernlochner}\ \emph {et~al.}(2018)\citenamefont
  {Bernlochner}, \citenamefont {Ligeti},\ and\ \citenamefont
  {Robinson}}]{bernlochner_dstst}%
  \BibitemOpen
  \bibfield  {author} {\bibinfo {author} {\bibfnamefont {F.~U.}\ \bibnamefont
  {Bernlochner}}, \bibinfo {author} {\bibfnamefont {Z.}~\bibnamefont
  {Ligeti}},\ and\ \bibinfo {author} {\bibfnamefont {D.~J.}\ \bibnamefont
  {Robinson}},\ }\bibfield  {title} {\bibinfo {title} {{Model-independent
  analysis of semileptonic $B$ decays to ${D}^{**}$ for arbitrary new
  physics}},\ }\href {https://doi.org/10.1103/PhysRevD.97.075011} {\bibfield
  {journal} {\bibinfo  {journal} {Phys. Rev. D}\ }\textbf {\bibinfo {volume}
  {97}},\ \bibinfo {pages} {075011} (\bibinfo {year} {2018})}\BibitemShut
  {NoStop}%
\bibitem [{\citenamefont {Bernlochner}\ and\ \citenamefont
  {Ligeti}(2017)}]{bernlochner_rdstst}%
  \BibitemOpen
  \bibfield  {author} {\bibinfo {author} {\bibfnamefont {F.~U.}\ \bibnamefont
  {Bernlochner}}\ and\ \bibinfo {author} {\bibfnamefont {Z.}~\bibnamefont
  {Ligeti}},\ }\bibfield  {title} {\bibinfo {title} {{Semileptonic ${B}_{(s)}$
  decays to excited charmed mesons with $e$, $\ensuremath{\mu}$,
  $\ensuremath{\tau}$ and searching for new physics with $R(D^{**})$}},\ }\href
  {https://doi.org/10.1103/PhysRevD.95.014022} {\bibfield  {journal} {\bibinfo
  {journal} {Phys. Rev. D}\ }\textbf {\bibinfo {volume} {95}},\ \bibinfo
  {pages} {014022} (\bibinfo {year} {2017})}\BibitemShut {NoStop}%
\bibitem [{\citenamefont {Bernlochner}\ \emph {et~al.}(2017)\citenamefont
  {Bernlochner} \emph {et~al.}}]{bernlochner_rd_rdst}%
  \BibitemOpen
  \bibfield  {author} {\bibinfo {author} {\bibfnamefont {F.~U.}\ \bibnamefont
  {Bernlochner}} \emph {et~al.},\ }\bibfield  {title} {\bibinfo {title}
  {{Combined analysis of semileptonic $B$ decays to $D$ and ${D}^{*}$:
  $R({D}^{(*)})$, $|{V}_{cb}|$, and new physics}},\ }\href
  {https://doi.org/10.1103/PhysRevD.95.115008} {\bibfield  {journal} {\bibinfo
  {journal} {Phys. Rev. D}\ }\textbf {\bibinfo {volume} {95}},\ \bibinfo
  {pages} {115008} (\bibinfo {year} {2017})},\ \bibinfo {note} {[Erratum: Phys.
  Rev. D \textbf{97}, 059902 (2018)]}\BibitemShut {NoStop}%
\bibitem [{\citenamefont {Ramirez}\ \emph {et~al.}(1990)\citenamefont
  {Ramirez}, \citenamefont {Donoghue},\ and\ \citenamefont
  {Burdman}}]{xu_hybrid}%
  \BibitemOpen
  \bibfield  {author} {\bibinfo {author} {\bibfnamefont {C.}~\bibnamefont
  {Ramirez}}, \bibinfo {author} {\bibfnamefont {J.~F.}\ \bibnamefont
  {Donoghue}},\ and\ \bibinfo {author} {\bibfnamefont {G.}~\bibnamefont
  {Burdman}},\ }\bibfield  {title} {\bibinfo {title} {{Semileptonic
  $b\ensuremath{\rightarrow}u$ decay}},\ }\href
  {https://doi.org/10.1103/PhysRevD.41.1496} {\bibfield  {journal} {\bibinfo
  {journal} {Phys. Rev. D}\ }\textbf {\bibinfo {volume} {41}},\ \bibinfo
  {pages} {1496} (\bibinfo {year} {1990})}\BibitemShut {NoStop}%
\bibitem [{\citenamefont {Ligeti}\ \emph {et~al.}(2022)\citenamefont {Ligeti},
  \citenamefont {Luke},\ and\ \citenamefont {Tackmann}}]{xutaunu}%
  \BibitemOpen
  \bibfield  {author} {\bibinfo {author} {\bibfnamefont {Z.}~\bibnamefont
  {Ligeti}}, \bibinfo {author} {\bibfnamefont {M.}~\bibnamefont {Luke}},\ and\
  \bibinfo {author} {\bibfnamefont {F.~J.}\ \bibnamefont {Tackmann}},\
  }\bibfield  {title} {\bibinfo {title} {{Theoretical predictions for inclusive
  $B\ensuremath{\rightarrow}{X}_{u}\ensuremath{\tau}\overline{\ensuremath{\nu}}$
  decay}},\ }\href {https://doi.org/10.1103/PhysRevD.105.073009} {\bibfield
  {journal} {\bibinfo  {journal} {Phys. Rev. D}\ }\textbf {\bibinfo {volume}
  {105}},\ \bibinfo {pages} {073009} (\bibinfo {year} {2022})}\BibitemShut
  {NoStop}%
\bibitem [{\citenamefont {Keck}\ \emph {et~al.}(2019)\citenamefont {Keck} \emph
  {et~al.}}]{fei}%
  \BibitemOpen
  \bibfield  {author} {\bibinfo {author} {\bibfnamefont {T.}~\bibnamefont
  {Keck}} \emph {et~al.},\ }\bibfield  {title} {\bibinfo {title} {{{The Full
  Event Interpretation}: {An exclusive tagging algorithm for the Belle II
  experiment}}},\ }\href {https://doi.org/10.1007/s41781-019-0021-8} {\bibfield
   {journal} {\bibinfo  {journal} {Comput. Softw. Big Sci.}\ }\textbf {\bibinfo
  {volume} {3}},\ \bibinfo {pages} {6} (\bibinfo {year} {2019})}\BibitemShut
  {NoStop}%
\bibitem [{\citenamefont {Aggarwal}\ \emph {et~al.}(2023)\citenamefont
  {Aggarwal} \emph {et~al.}}]{Remu}%
  \BibitemOpen
  \bibfield  {author} {\bibinfo {author} {\bibfnamefont {L.}~\bibnamefont
  {Aggarwal}} \emph {et~al.} (\bibinfo {collaboration} {Belle II
  Collaboration}),\ }\bibfield  {title} {\bibinfo {title} {{Test of
  light-lepton universality in the rates of inclusive semileptonic $B$-meson
  decays at Belle II}},\ }\href
  {https://doi.org/10.1103/PhysRevLett.131.051804} {\bibfield  {journal}
  {\bibinfo  {journal} {Phys. Rev. Lett.}\ }\textbf {\bibinfo {volume} {131}},\
  \bibinfo {pages} {051804} (\bibinfo {year} {2023})}\BibitemShut {NoStop}%
\bibitem [{\citenamefont {Milesi}\ \emph {et~al.}(2020)\citenamefont {Milesi},
  \citenamefont {Tan},\ and\ \citenamefont {Urquijo}}]{LeptonIDBDT}%
  \BibitemOpen
  \bibfield  {author} {\bibinfo {author} {\bibfnamefont {M.}~\bibnamefont
  {Milesi}}, \bibinfo {author} {\bibfnamefont {J.}~\bibnamefont {Tan}},\ and\
  \bibinfo {author} {\bibfnamefont {P.}~\bibnamefont {Urquijo}},\ }\bibfield
  {title} {\bibinfo {title} {{Lepton identification in Belle II using
  observables from the electromagnetic calorimeter and precision trackers}},\
  }\href {https://doi.org/10.1051/epjconf/202024506023} {\bibfield  {journal}
  {\bibinfo  {journal} {EPJ Web Conf.}\ }\textbf {\bibinfo {volume} {245}},\
  \bibinfo {pages} {06023} (\bibinfo {year} {2020})}\BibitemShut {NoStop}%
\bibitem [{\citenamefont {Amhis}\ \emph {et~al.}(2023)\citenamefont {Amhis}
  \emph {et~al.}}]{hflav}%
  \BibitemOpen
  \bibfield  {author} {\bibinfo {author} {\bibfnamefont {Y.~S.}\ \bibnamefont
  {Amhis}} \emph {et~al.} (\bibinfo {collaboration} {Heavy Flavor Averaging
  Group Collaboration}),\ }\bibfield  {title} {\bibinfo {title} {{Averages of
  $b$-hadron, $c$-hadron, and $\tau$-lepton properties as of 2021}},\ }\href
  {https://doi.org/10.1103/PhysRevD.107.052008} {\bibfield  {journal} {\bibinfo
   {journal} {Phys. Rev. D}\ }\textbf {\bibinfo {volume} {107}},\ \bibinfo
  {pages} {052008} (\bibinfo {year} {2023})}\BibitemShut {NoStop}%
\bibitem [{\citenamefont {Lees}\ \emph {et~al.}(2016)\citenamefont {Lees} \emph
  {et~al.}}]{dpipi_babar}%
  \BibitemOpen
  \bibfield  {author} {\bibinfo {author} {\bibfnamefont {J.~P.}\ \bibnamefont
  {Lees}} \emph {et~al.} (\bibinfo {collaboration} {BaBar Collaboration}),\
  }\bibfield  {title} {\bibinfo {title} {{Observation of
  $\Bbar\ensuremath{\rightarrow}{D}^{(*)}\text{
  }{\ensuremath{\pi}}^{+}{\ensuremath{\pi}}^{\ensuremath{-}}{\ensuremath{\ell}}^{\ensuremath{-}}\overline{\ensuremath{\nu}}$
  decays in ${e}^{+}{e}^{\ensuremath{-}}$ collisions at the
  $\mathrm{\ensuremath{\Upsilon}}(4S)$ resonance}},\ }\href
  {https://doi.org/10.1103/PhysRevLett.116.041801} {\bibfield  {journal}
  {\bibinfo  {journal} {Phys. Rev. Lett.}\ }\textbf {\bibinfo {volume} {116}},\
  \bibinfo {pages} {041801} (\bibinfo {year} {2016})}\BibitemShut {NoStop}%
\bibitem [{\citenamefont {Bernlochner}\ \emph {et~al.}(2020)\citenamefont
  {Bernlochner} \emph {et~al.}}]{hammer}%
  \BibitemOpen
  \bibfield  {author} {\bibinfo {author} {\bibfnamefont {F.~U.}\ \bibnamefont
  {Bernlochner}} \emph {et~al.},\ }\bibfield  {title} {\bibinfo {title} {{Das
  ist der {HAMMER}: Consistent new physics interpretations of semileptonic
  decays}},\ }\href {https://doi.org/10.1140/epjc/s10052-020-8304-0} {\bibfield
   {journal} {\bibinfo  {journal} {Eur. Phys. J. C}\ }\textbf {\bibinfo
  {volume} {80}},\ \bibinfo {pages} {883} (\bibinfo {year} {2020})}\BibitemShut
  {NoStop}%
\bibitem [{\citenamefont {Boyd}\ \emph {et~al.}(1995)\citenamefont {Boyd},
  \citenamefont {Grinstein},\ and\ \citenamefont {Lebed}}]{BGL}%
  \BibitemOpen
  \bibfield  {author} {\bibinfo {author} {\bibfnamefont {C.~G.}\ \bibnamefont
  {Boyd}}, \bibinfo {author} {\bibfnamefont {B.}~\bibnamefont {Grinstein}},\
  and\ \bibinfo {author} {\bibfnamefont {R.~F.}\ \bibnamefont {Lebed}},\
  }\bibfield  {title} {\bibinfo {title} {{{Constraints on form-factors for
  exclusive semileptonic heavy to light meson decays}}},\ }\href
  {https://doi.org/10.1103/PhysRevLett.74.4603} {\bibfield  {journal} {\bibinfo
   {journal} {Phys. Rev. Lett.}\ }\textbf {\bibinfo {volume} {74}},\ \bibinfo
  {pages} {4603} (\bibinfo {year} {1995})}\BibitemShut {NoStop}%
\bibitem [{\citenamefont {Glattauer}\ \emph {et~al.}(2016)\citenamefont
  {Glattauer} \emph {et~al.}}]{DlnuBGL}%
  \BibitemOpen
  \bibfield  {author} {\bibinfo {author} {\bibfnamefont {R.}~\bibnamefont
  {Glattauer}} \emph {et~al.} (\bibinfo {collaboration} {Belle
  Collaboration}),\ }\bibfield  {title} {\bibinfo {title} {{Measurement of the
  decay
  $B\ensuremath{\rightarrow}D\ensuremath{\ell}{\ensuremath{\nu}}_{\ensuremath{\ell}}$
  in fully reconstructed events and determination of the
  Cabibbo-Kobayashi-Maskawa matrix element $|{V}_{cb}|$}},\ }\href
  {https://doi.org/10.1103/PhysRevD.93.032006} {\bibfield  {journal} {\bibinfo
  {journal} {Phys. Rev. D}\ }\textbf {\bibinfo {volume} {93}},\ \bibinfo
  {pages} {032006} (\bibinfo {year} {2016})}\BibitemShut {NoStop}%
\bibitem [{\citenamefont {Ferlewicz}\ \emph {et~al.}(2021)\citenamefont
  {Ferlewicz}, \citenamefont {Urquijo},\ and\ \citenamefont
  {Waheed}}]{DstlnuBGL}%
  \BibitemOpen
  \bibfield  {author} {\bibinfo {author} {\bibfnamefont {D.}~\bibnamefont
  {Ferlewicz}}, \bibinfo {author} {\bibfnamefont {P.}~\bibnamefont {Urquijo}},\
  and\ \bibinfo {author} {\bibfnamefont {E.}~\bibnamefont {Waheed}},\
  }\bibfield  {title} {\bibinfo {title} {{Revisiting fits to $B^0\to
  D^{*-}\ell^{+}\nu_\ell$ to measure $|V_\text{cb}|$ with novel methods and
  preliminary LQCD data at nonzero recoil}},\ }\href
  {https://doi.org/10.1103/PhysRevD.103.073005} {\bibfield  {journal} {\bibinfo
   {journal} {Phys. Rev. D}\ }\textbf {\bibinfo {volume} {103}},\ \bibinfo
  {pages} {073005} (\bibinfo {year} {2021})}\BibitemShut {NoStop}%
\bibitem [{\citenamefont {Caprini}\ \emph {et~al.}(1998)\citenamefont
  {Caprini}, \citenamefont {Lellouch},\ and\ \citenamefont {Neubert}}]{CLN}%
  \BibitemOpen
  \bibfield  {author} {\bibinfo {author} {\bibfnamefont {I.}~\bibnamefont
  {Caprini}}, \bibinfo {author} {\bibfnamefont {L.}~\bibnamefont {Lellouch}},\
  and\ \bibinfo {author} {\bibfnamefont {M.}~\bibnamefont {Neubert}},\
  }\bibfield  {title} {\bibinfo {title} {{Dispersive bounds on the shape of $B
  \to D^{(*)}\ell \nu $ form factors}},\ }\href
  {https://doi.org/10.1016/S0550-3213(98)00350-2} {\bibfield  {journal}
  {\bibinfo  {journal} {Nucl. Phys. B}\ }\textbf {\bibinfo {volume} {530}},\
  \bibinfo {pages} {153} (\bibinfo {year} {1998})}\BibitemShut {NoStop}%
\end{thebibliography}%

\clearpage
\section{Supplemental material}

\subsection{Validation of the simulation reweighting}

\begin{figure*}[h!]
	\centering
	\begin{subfigure}[b]{.98\linewidth}
        \includegraphics[width=\textwidth]{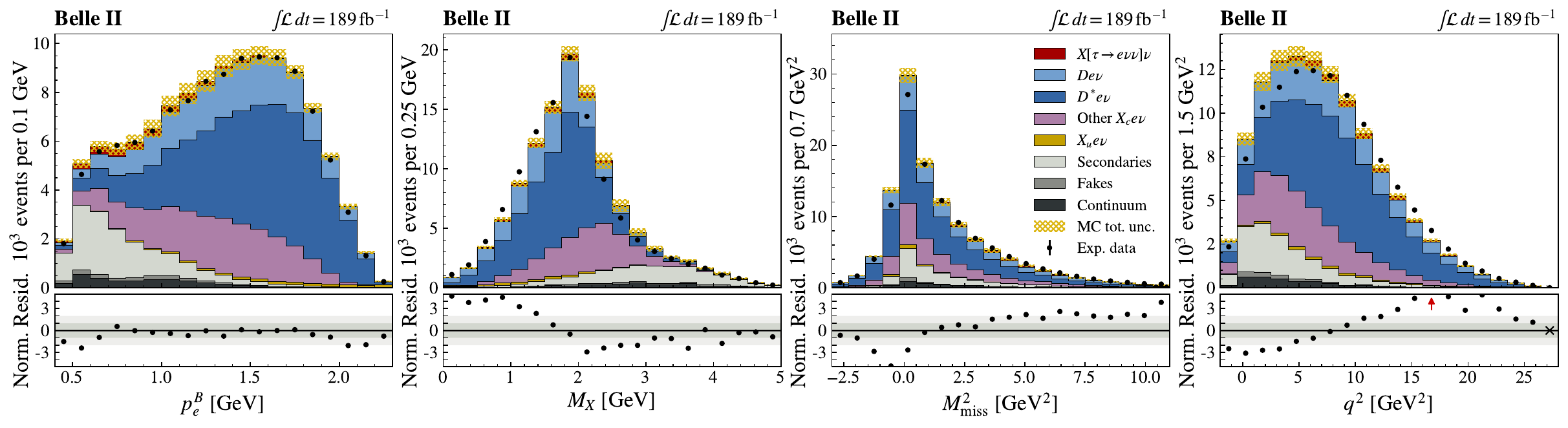}
        \includegraphics[width=\textwidth]{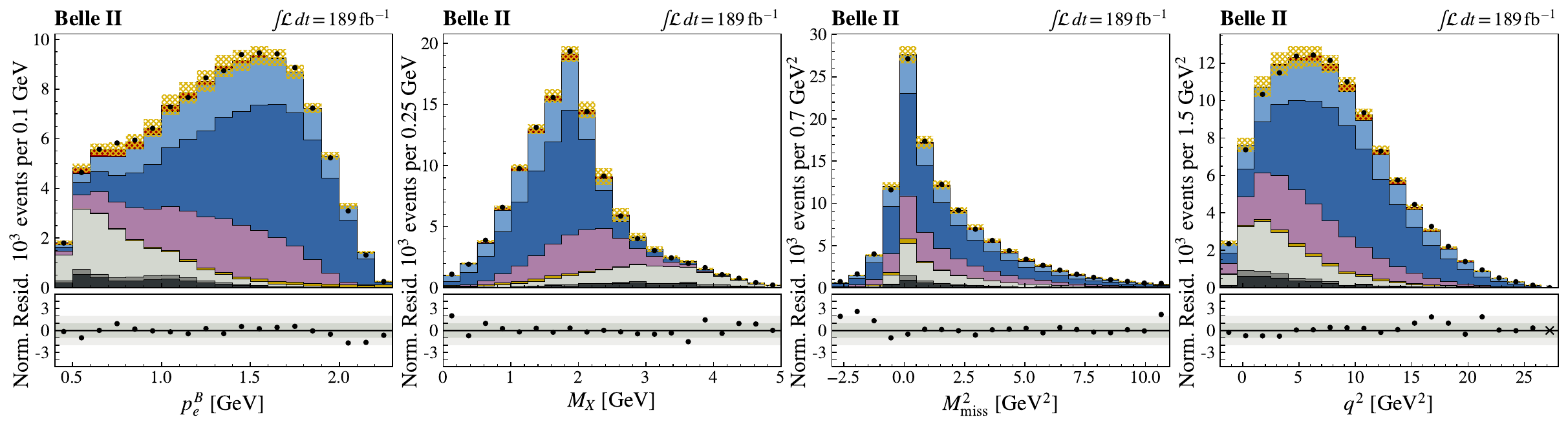}
		\caption{Electron channel before (top) and after (bottom) the simulation reweighting.}
	\end{subfigure}
	\begin{subfigure}[b]{.98\linewidth}
        \includegraphics[width=\textwidth]{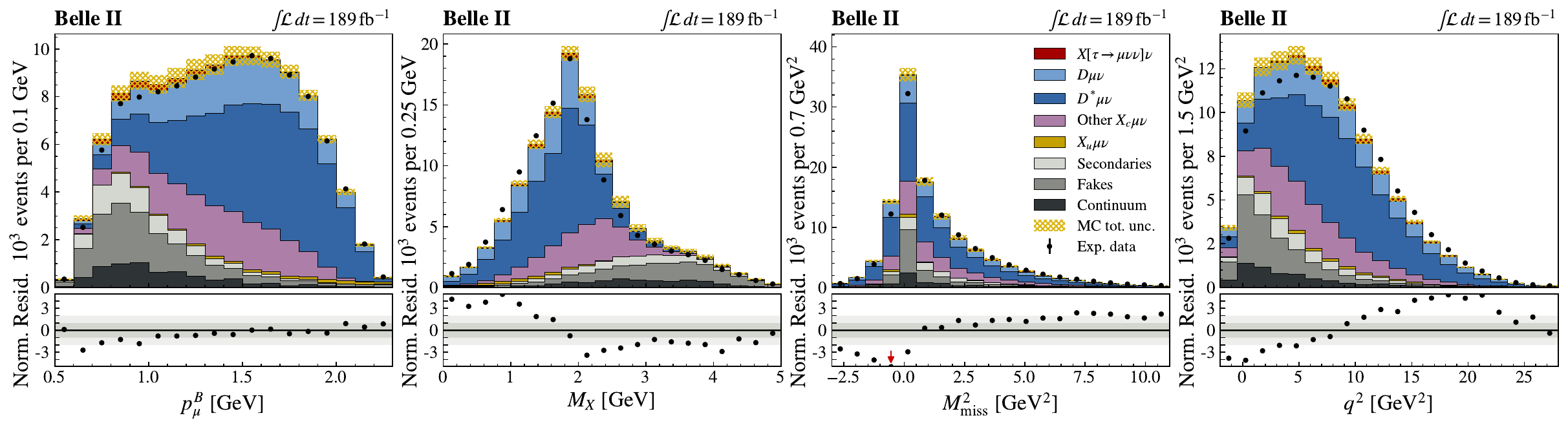}
        \includegraphics[width=\textwidth]{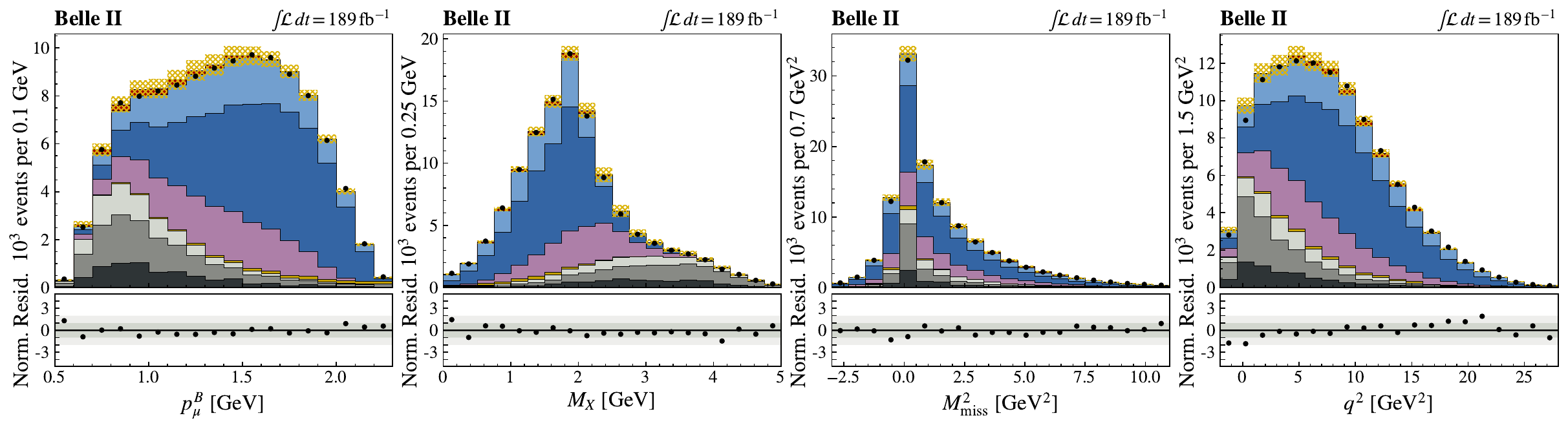}
		\caption{Muon channel before (top) and after (bottom) the simulation reweighting.}
	\end{subfigure}
	\caption{The effect of the \mx and $(\pllab - \mx)$-based reweighting on four key kinematic quantities for the electron (a, top) and muon (b, bottom) channels. The top rows show the pre-reweighting distributions in simulated (filled histograms) and experimental (black points) data, with their uncertainty-normalized disagreement (``Norm. Resid.'') shown below. The bottom row of plots shows the reweighted distributions, with significantly reduced residuals.}
	\label{fig:reshaping_effects_on_kinematic_vars}
\end{figure*}

In Fig.~\ref{fig:reshaping_effects_on_kinematic_vars} we illustrate the effect of the \mx and $(\pllab, \mx)$-based reweightings of \xlnu and \BB background events on four key kinematic quantities, including the signal extraction quantities \plB and \mmsq, the calibration quantity \mx, and a control quantity $\qsq=((\sqrt{s}, \vec{0}) -P_{\btag}^{\cms} -P_{X}^{\cms} )^2$ that is used neither in the signal extraction nor in the reweighting. The reweightings improve the agreement between experimental and simulated data, best quantified by the normalized residuals, defined as the difference between simulated and experimental yields divided by the quadrature sum of their statistical uncertainties. 

The reweightings have negligible effect on the \plB shape for \bxclnu events as the lepton momentum is largely independent of the hadronic $X$ system. For \BB backgrounds, the lepton candidate's role in the $B$-meson decay chain is more complex, and the shape of the component is modified in the reweighting in the same-flavor control sample. This is reflected in modest improvements to the agreement in the low-\plB region after the reweighting.\par

The remaining three quantities, \mx, \mmsq, and \qsq, depend directly on the calibration quantity \mx. We observe large and simultaneous improvements in all of these quantities, seen in the figure as reduced residuals across all bins. The major improvements in \mmsq and \qsq, in particular, suggest that the reweightings mitigate the underlying modeling errors. The multiplicities of charged kaons, charged pions, and photons, not shown, also improve after reweighting. We tested the impact in various signal-depleted control regions in addition to the two mentioned: high-\mx ($\mx > 3\,\gevcc$, $79\%$ background and continuum events) and low-\mmsq ($\mmsq < 1.5\,\gevsqcccc$, $78\%$ \bxlnu events, $21\%$ background and continuum events). We observe similar improvements, supporting the validity of the reweightings in all kinematic regions.

\subsection{Relationship between \RX and \RDorDst}

In order to understand how the measured tau-to-light-lepton ratio of inclusive $B$-meson branching fractions $\RX_\text{exp}$ relates to the \RDorDst anomalies, it is essential to control for the variety of additional decays included in \bxtaulnu. Aside from the $D$ and $D^{*}$ hadrons that are selected in the exclusive \RDorDst measurements, our events also contain $D^{**}$, nonresonant $X_c$ ($D_\text{gap}$), and minimal $X_u$ contributions. 

For this purpose, the measured light-lepton branching fractions \BFdlnu, \BFdstlnu, and \BFxlnu are needed. For the inclusive branching fraction, we use the latest value from Ref.~\cite{hflav},
\begin{equation}
    \BFxlnu = (10.84 \pm 0.16)\%.
\end{equation}
For the exclusive branching fractions, we calculate the isospin-averaged values of the $B^+$ and $B^0$ measurements summarized in the same reference, and use the arithmetic mean of the different $B$-meson lifetimes to obtain
\begin{align}
    \BFdlnu &=  (2.27 \pm 0.06)\% \\
    \BFdstlnu &=  (5.23 \pm 0.10)\%.
\end{align}

The standard model prediction of \RXc is calculated in Ref.~\cite{kvos}, superseding Ref.~\cite{mannel_rxc}, and in a different scheme in Ref.~\cite{ligeti_rxc}, while Refs.~\cite{kvos, xutaunu} provide theoretical input for either fully inclusive \RX or \RXu. We average the predicted \RX value of Ref.~\cite{kvos} with a combination of \RXc and \RXu from Refs.~\cite{ligeti_rxc, xutaunu} using inclusive branching fractions from Ref.~\cite{hflav} to derive 
\begin{equation}
    \RX_\text{SM} = 0.223 \pm 0.005
\end{equation}
and accordingly
\begin{equation}\label{eq:bf_xtaunu}
    \BFxtaunu = (2.42 \pm 0.06)\% \text{.}
\end{equation}

On its own, $\RX_\text{exp}$ imposes an upper bound on the sum of measured \RD and \RDst values,
\begin{align}\label{eq:rx_ge_rdordst}
    \RX_\text{exp} \times \BFxlnu = \BFxtaunu &\ge \notag \\
    \BFdtaunu + \BFdsttaunu &= \notag\\
    \RD \times \BFdlnu + \RDst \times \BFdstlnu.
\end{align}
By inserting the expected additional contributions to \xtaunu we can write
\begin{align}\label{eq:rx_eq_rdordst}
\BFxtaunu &= \BFdtaunu + \BFdsttaunu \notag \\
&+ \BFdgapxutaunu.
\end{align}
The size of these additional contributions can be calculated in the standard model by inserting predicted values for the semitauonic branching fractions based on measured light-lepton branching fractions,
\begin{align}\label{eq:bf_dstst_etc}
    \BFdgapxutaunu_\text{SM} &= \RX_\text{SM} \times \BFxlnu\notag \\
    &- \RD_\text{SM} \times \BFdlnu\notag \\
    &- \RDst_\text{SM} \times \BFdstlnu \notag \\
    &= (0.41\pm0.08)\%.
\end{align}
This corresponds to $(17.1 \pm 2.8)\%$ of the total semitauonic branching fraction given in Eq.~\eqref{eq:bf_xtaunu}.
By assuming that all unmeasured additional semitauonic contributions are standard-model-like, we define the reduced ratio $\RX^\dagger$ as 
\begin{align}\label{eq:rxdagger}
\RX^\dagger &\equiv \frac{\BFxtaunu - \BFdgapxutaunu_\text{SM}}{\BFxlnu} \notag \\
&= \RX_\text{exp} - \frac{\BFdgapxutaunu_\text{SM}}{\BFxlnu}
\end{align}
so that the full constraining power of $\RX_\text{exp}$ on \RDorDst can be expressed as
\begin{align}\label{rx_on_rdordst_plane}
\RX^\dagger \times \BFxlnu = \notag \\
x_{\RD} \times \BFdlnu + y_{\RDst} \times  \BFdstlnu \text{.}
\end{align}
Here, we have replaced the experimental value of \RD (\RDst) by the running quantity $x_{\RD}$ ($y_{\RDst}$) so that these findings can directly be summarized in the $\RD-\RDst$ plane that compares measured and predicted \RDorDst values. Solving Eq.~\eqref{rx_on_rdordst_plane} for $y_{\RDst}$ converts the measured \RX value into a straight line on the plane as depicted in Fig.~\ref{fig:result_plot}.

\end{document}